\documentclass[11pt,american]{article}

\usepackage{amsmath}
\usepackage{amstext}
\usepackage{amssymb}
\usepackage{array}
\usepackage{babel}
\usepackage{booktabs}
\usepackage{cancel}
\usepackage{comment}
\usepackage{eurosym}
\usepackage[T1]{fontenc}
\usepackage[a4paper]{geometry}
\usepackage{graphicx}
\usepackage[unicode=true,pdfusetitle,bookmarks=true,bookmarksnumbered=false,bookmarksopen=false,breaklinks=false,pdfborder={0 0 0},backref=false,colorlinks=false]{hyperref}
\usepackage[utf8]{inputenc}
\usepackage{lipsum}
\usepackage{lmodern}

\usepackage{mathdots}
\usepackage{mathtools}
\usepackage{pdfcolmk}
\usepackage{rotating}
\usepackage{setspace}
\usepackage{stackrel}
\usepackage{subfig}
\usepackage{subscript}
\usepackage{tabularx} 
\usepackage{textcomp}
\usepackage{units}
\usepackage{url}
\usepackage{floatrow}
\usepackage{verbatim}
\usepackage[dvipsnames]{xcolor}

\usepackage{cleveref}

\usepackage[backend=biber, citestyle=numeric, bibstyle=numeric, sorting=none, natbib, maxbibnames=99]{biblatex}
\newcommand{\incidence}[1][n]{K_{#1,\ell}}
\newcommand{\ptdf}[1][n]{H_{\ell,#1}}
\newcommand{\flow}{f_{\ell,t}}
\newcommand{\reactivepower}{q_{n,t}}
\newcommand{\injection}[1][n]{p_{#1,t}}

\newcommand{\allocatePeer}[1][n \rightarrow m]{A_{#1,t}}
\newcommand{\allocateFlow}[1][n]{F_{#1,\ell,t}}
\newcommand{\allocateInjection}{P_{n,m,t}}
\newcommand{\allocateTransaction}[1][n \rightarrow m]{A_{#1,\ell,t}}
\newcommand{\Forall}[1]{\hspace{20pt} \forall \,\, #1}

\newcommand{\incidenceM}{K}
\newcommand{\flowM}{f}
\newcommand{\injectionM}{p}

\newcommand{\DirectedIncidence}{\mathcal{K}}
\newcommand{\InverseAPInjection}{\mathcal{J}}
\newcommand\diag[1]{\operatorname{diag}\left(#1\right)}
\newcommand{\netconsumption}[1][n]{p^{-}_{#1,t}}
\newcommand{\netproduction}[1][n]{p^{+}_{#1,t}}
\newcommand{\selfconsumption}[1][n]{p^{\circ}_{#1,t}}


\pdfpageheight\paperheight
\pdfpagewidth\paperwidth

\graphicspath{{Figures-update/}}

\begin{document}


\title{Techno-economic criteria to evaluate power flow allocation schemes}

\author{
Fabian Hofmann\thanks{Frankfurt Institute for Advanced Studies (FIAS), Ruth-Moufang-Straße 1, 60438 Frankfurt am Main, Germany
}\footnotemark[1], 
Alexander Zerrahn\thanks{DIW Berlin, Mohrenstraße 58, 10117 Berlin, Germany} \thanks{Corresponding author, azerrahn@diw.de},
Carlos Gaete-Morales\footnotemark[2]
}

\date{}

\maketitle
\begin{center}
\vspace{-0.6cm}

\par\end{center}


\begin{center}
\vspace{0.6cm}

\par\end{center}
\begin{abstract}
We develop novel quantitative techno-economic evaluation criteria for power flow allocation schemes. Such schemes assign which nodes are responsible for which proportion of power flows on a line in a meshed electricity transmission network. As this allocation is, as such, indeterminate, the literature has proposed a number of dedicated schemes. To better understand the implications of applying different schemes, our criteria comprise their (i) fairness, (ii) plausibility, (iii) uniqueness, and (iv) stability. We apply, illustrate, and discuss these criteria for four prominent schemes based on results of a detailed electricity sector model with linear power flow for a German mid-term future case.

\end{abstract}

\noindent 
\vspace{5cm}

\textit{Keywords}: linear power flow model, flow allocation, power flow tracing, electricity network \thispagestyle{empty}

\noindent \newpage{}
\setcounter{page}{1}


\section{Introduction}\label{sec:intro}

Power flows within an alternating current (AC) network cannot be scheduled to specific lines, but spread according to physical laws. For a given network topology and nodal net injections, Kirchhoff's laws determine the flow on all lines, but it is \textit{ex ante} indeterminate which proportion of the flow on this line is due to the injection or withdrawal of each node. To this end, researchers have developed a range of dedicated flow allocation schemes. Such schemes allocate the physical electricity flow on a line to nodes in the network. In general, different flow allocation schemes lead, by construction, to different allocations of flows to nodes. 

The literature so far applies these schemes largely in isolation. As such, a researcher or analyst is free to pick among several alternatives. Yet the resulting allocations differ and with them the implications for interpreting the network situation and potential technical or economic measures. For instance, one scheme may attribute a large proportion of flows to nodes close to that line while another scheme attributes flows to nodes farther away; or, one scheme rather may attribute flows to nodes that depend on power exports while another scheme attributes more flows to transit nodes. Therefore, a deeper understanding of economic network use and management requires a systematic comparison and evaluation of flow allocation schemes. Yet such systematic evaluation is missing.

We fill this knowledge gap. We contribute to the literature by designing novel quantitative criteria to evaluate flow allocation schemes. Specifically, we develop techno-economic metrics capturing the $(i)$ fairness, $(ii)$ plausibility, $(iii)$ uniqueness, and $(iv)$ stability of flow allocations resulting from different schemes. To illustrate our criteria, we apply these metrics to four prominent flow allocation schemes for a sample network for a German future scenario and discuss the results in detail. For transparency, reproducibility, and as a service to the profession, we make all analyzed allocation schemes available as software packages under a permissive license.

Our results inform researchers and policy analysts concerned with modeling, managing, and regulating electricity networks. The electricity network assumes a central mediating role to connect generation and consumption. This gains special relevance as power systems are changing fast around the world. On the demand side, new consumers emerge like, for instance, from mobility and heating; on the supply side, generation technologies see a shift toward renewable energy supply. Network management can be facilitated when network use by generation and consumption agents at different nodes can be determined. This applies to the short term, when trade positions or congestion are managed, and the long term, when lines get expanded. 

Our paper is most closely related to the literature on flow allocation that started with Bialek's flow tracing~\citep{bialek_tracing_1996}, leading to Average Participation flow allocation, and Marginal Participation~\citep{rudnick_marginal_1995}. Subsequently, more methods have been developed such as Equivalent Bilateral Exchanges~\citep{galiana_transmission_2003}, Z-bus allocation~\citep{conejo_z-bus_2001,conejo_z-bus_2007}, and others. The schemes have been put to application by studying cases for test systems~\citep{tranberg_flow-based_2018} or the European electricity network~\citep{horsch_flow_2018,brown_transmission_2015}. While the mathematical foundations of the four mentioned approaches are laid out and applications exist, we provide a comparative overview, illustration, and evaluation according to newly developed quantitative criteria. 

The remainder of this paper proceeds as follows. Section~\ref{sec:basics} summarizes linear power flow basics and characterizes different power flow allocation schemes. Section~\ref{sec:application} puts the schemes to application and compares as well as characterizes the emerging allocations. Section~\ref{sec:eval} devises quantitative evaluation criteria for the allocation schemes and discusses economic, policy, and practical consequences for their application. Section~\ref{sec:eval_acdc} provides a robustness check of our criteria concerning the linear load flow approximation of AC power flows. Section~\ref{sec:conclusion} concludes and outlines avenues for future research.

\section{Linear power flow and power flow allocation}\label{sec:basics}


\subsection{Linear power flow}\label{sec:basics_subsec:dclf}
\begin{subequations}
Consider an electricity network with~$N$ nodes and~$L$ lines. The power flow on each line~$\flow$ is a function of the active nodal power injection~$\injection$ and the reactive nodal power injection~$\reactivepower$. Calculating the full alternating current flow for large networks constitutes a non-linear problem that requires elaborate and computationally demanding algorithms such as Newton-Raphson, Decoupled Load Flow or Holomorphic Embedding~\citep{trias_holomorphic_2012}. Yet in many applications in system planning it proves to be convenient to represent the power flow through a linear approximation. This linear power flow approach, also referred to as DC load flow approach, goes back to~\citet{schweppe_1988} and has become a quasi standard in much of applied research. It allows to embed the electricity network in broader analyses that rely on (mixed integer) linear optimization like capacity expansion planning~\citep{horsch_pypsa-eur_2018,leuthold_2012} or optimal dispatch~\citep{kunz_congestion_2016,peker_benefits_2018,qadrdan_benefits_2017}.

The linear power flow approach assumes that all nodal voltage magnitudes are equal, and the series resistances are small compared to series reactances. In addition, voltage angle differences across a line are assumed to be small, and no shunt admittances (at nodes or series) to ground are present. Under these assumptions, reactive power~$\reactivepower$ can be neglected, and the sine representation of voltage angles differences between each node and the slack node be linearly approximated with a sufficiently small error.

These assumptions mainly suit the power flow on the transmission system level, where high voltages and small resistances lead to a flow which is mostly driven by active power injection~$\injection$. As a result of this approximation, the Power Transfer Distribution Factors (PTDF)~$\ptdf$ map the active power injection~$\injection$ to the active power flow~$\flow$:
\begin{align}	
\flow &= \sum_n \, \ptdf \, \injection \Forall{\ell,t}\label{eq:powerflow_equation_lin}
\end{align}

The PTDF have a degree of freedom that can be used to define which node or which set of nodes absorb power imbalances in case the power injection~$\sum_n \injection$ is not balanced. This node or these nodes are referred to as slack node(s). To end this, the nodal power injection balances out the sum of incoming and outgoing flows:
\begin{align}
\injection = \sum_\ell \incidence \, \flow \Forall{n,t},\label{eq:nodal_balance_lin}
\end{align}

where~$\incidence$ denotes the elements of the incidence matrix~$\incidenceM$. Entry~$\incidence$ is equal to~$1$ if line~$\ell$ starts at node~$n$, equal to~$-1$ if~$\ell$ ends at node~$n$, and zero otherwise. Equation~\eqref{eq:nodal_balance_lin} reflects the power flow conservation law at each node that all power injected at a node~$n$ must flow into the network, if positive, or be withdrawn from the network, if negative.

\end{subequations}

\subsection{Power flow allocation schemes}\label{sec:basics_subsec:alloc}
\begin{subequations}\label{eq:alloc_basics}
The central equation~\eqref{eq:powerflow_equation_lin} specifies the relationship between the power injection at all nodes and the flows on all lines. Yet it is, as such, silent about the specific contribution how strongly an injection or withdrawal at a node contributes to flows on a line. Deriving these contributions requires assumptions and dedicated methods which nodes serve each other with power and using which line, referred to as power flow allocation schemes. 

Figure~\ref{fig:flow_decomposition} gives a stylized example of the problem in a three-node network. The panel on the left-hand side shows the nodal net injections, proportional to the areas of the circles, and the resulting (linearized) physical power flows~$\flow$ as arrows with the according proportional size. The two lower nodes sources without demand; the upper node is a sink without supply. The smaller panels on the right-hand side panel show two possible allocation schemes. 

\begin{figure}[b!]
\centering
\includegraphics[width=\linewidth]{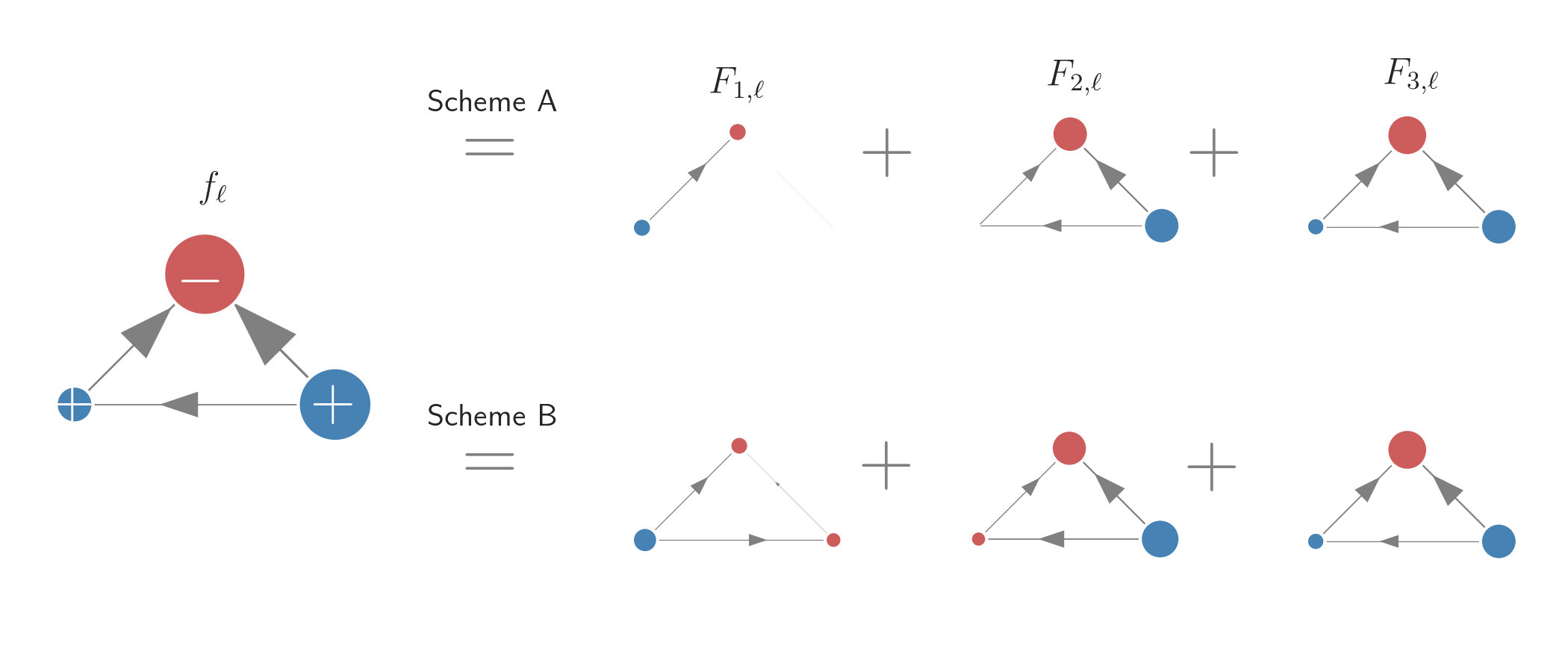}
\caption{Each power flow~$f_{\ell}$ can be decomposed into nodal contributions~$F_{n,\ell}$. However, the decomposition is non-unique and needs assumptions on how the power between producers and consumers is allocated.}
\label{fig:flow_decomposition}
\end{figure}

According to scheme~$A$, node~$1$ in the lower left corner only gets allocated a flow~$F_{1,\ell}$ on the line that connects it with the sink. Node~$2$ in the lower right corner gets allocated the flow on the line that connects it with the sink, the flow on the lower line, and part of the flow connecting node~$1$ with the sink. The upper sink node~$3$ gets allocated a half of the flows on all lines. According to an alternative scheme~$B$, the lower left node~$1$ additionally gets allocated a flow on the lower line that runs opposite to the physical flow direction and a small proportion of the flow on the line that connects the lower right node to the sink. Accordingly, the lower right node~$2$ gets allocated a larger flow on the lower line in the physical flow direction and a somewhat smaller flow on the line that connects it to the sink than under scheme~$A$. Adding the subflows on all lines yields the same total flow; yet the allocation to the nodes is different. Likewise, it is also indeterminate which proportion of an allocated subflow on a line is attributed to a supplying source and a receiving sink. As such, there is no natural mechanism or criterion which allocation scheme should be superior or preferred.

If a specific allocation scheme is picked, the power flow can be considered as a linear combination of nodal contributions. To this end, different approaches specify different assumptions how to allocate flows to nodes. As a key result, an allocation scheme specifies an allocated subflow~$\allocateFlow$ induced by node~$n$ on line~$\ell$ at time step~$t$. That is, whether~$n$ injects or extracts power from the network contributes to the flow on each line~$\ell$. The sum of all allocated subflows must result in the actual power flow:
\begin{align}
\flow &= \sum_n \allocateFlow  \Forall{\ell},t 
\label{eq:alloc_flow_sum}
\end{align} 
Inserting this relation into~\eqref{eq:nodal_balance_lin} leads to allocated injections, also referred to as \textit{virtual injection patterns}, that relate the net or gross injection of each node to all other nodes:
\begin{align}
\injection &= \sum_\ell \incidence \,\left( \sum_m \allocateFlow[m] \right) \\
&= \sum_m \allocateInjection  \Forall{n,t}
\label{eq:alloc_injection_sum}
\end{align}
The value~$\allocateInjection$ indicates the power injections at node~$m$ caused by node~$n$. If, for example, node~$n$ is a net consumer, other nodes must come up with net positive injections providing power for~$n$. The amount of power provided by~$m$ for~$n$ is given by the virtual injection patterns~$\allocateInjection$ that, in turn, are derived from different flow allocation schemes. Based on a complete specification of~$\allocateInjection$ that also specifies the split between the source~$n$ and the sink~$m$, the allocated flows on all lines~$\allocateFlow$ can be determined according to 
\begin{align}
\allocateFlow = \sum_m \ptdf[m] \, \allocateInjection \Forall{\ell,n,t}
\label{eq:alloc_vip_to_allocflow}
\end{align}

In order to allocate flows on the basis of given virtual injection patterns in transport networks or networks across multiple synchronous zones, the PTDF~$\ptdf$ can be adjusted according to the findings in~\citep{hofmann_flow_2020}.

Alternatively, an allocation scheme may be based on a defined power transaction, $\allocateTransaction$, also referred to as peer-to-peer relation. It defines the power that node~$n$ supplies to node~$m$ in time step~$t$, inducing subflow on line~$\ell$. As such, there is a degree of freedom whether the supplying or receiving node is responsible for that subflow. A straightforward, and common, proposition is to weigh the contributions in a fifty-fifty manner, which, in mathematical terms, leads to:
\begin{align}
\allocateFlow = \dfrac{1}{2} \sum_m \left( \allocateTransaction +\allocateTransaction[m \rightarrow n] \right) \Forall{\ell,n,t} 
\label{eq:transactions_to_flow_allocations}
\end{align}

Different flow allocation schemes start out from either specifying virtual injection patterns or peer-to-peer relations. Yet both approaches can be transferred into each other. After this short presentation of the most important quantities and relations, we discuss four relevant flow allocation schemes. In doing so, we restrict ourselves to the practically relevant linear flow approximation of meshed AC networks. For transparency, reproducibility~\citep{pfenninger_2017}, and as a service the research community, we also provide the Python package~\emph{netallocation} \cite{netallocation_package} that implements the four discussed schemes for linear power sector models.\footnote{The package~netallocation is available under~\url{https://github.com/FRESNA/netallocation}.} 

\end{subequations}

\subsubsection*{Average Participation}\label{sec:alloc_subsec:alloc_subsubsec:ap}
The Average Participation ($AP$) allocation scheme was first introduced by~\citet{kirschen_contributions_1997} and~\citet{bialek_tracing_1996}. The basic concept applies the principle of proportional sharing to trace a power flow from source to sink (downstream) or from sink to source (upstream). 

For downstream tracing, the scheme takes into account all lines that transport power away from a node~$n$ with positive net injection. In this regard, the partial flow assigned to~$n$, $\allocateFlow$, splits proportionally to the amount of power that flows on these lines. When~$\allocateFlow$ meets the next node~$m$, it is either (partially) absorbed if~$m$ is a sink or it mixes with partial flows that originate from~$m$ if~$m$ is a source. If~$m$ has outgoing flows, $\allocateFlow$ again splits proportionally to those. In that way, partial flows originating from all sources are traced through the network to their eventual sinks, rendering the $AP$ scheme a full peer-to-peer transaction~$\allocateTransaction$.

Analogously for upstream tracing, the scheme takes into account all lines with incoming flows to a node~$n$ with negative net injection. All imports to~$n$ are allocated among lines that transport power to~$n$ proportional to the total flow on that lines. An adjacent node~$m$ that exports to node~$n$ either is a sink or a source. If it is a sink, no flow from~$m$ to~$n$ is allocated to that line. If it is a source, the partial flow on the connecting line~$\ell$ originates (partially) from~$m$. If~$m$ also has incoming flows, flow~$\allocateFlow$ that does not originate from~$m$, is assumed to originate from the nodes adjacent to~$m$ proportional to the total flows on those lines. Thus, the algorithm traces flows backwards through the network until all partial flows ending in sinks have a defined starting point, and a full peer-to-peer transaction~$\allocateTransaction$ is specified. 

Based on these transactions, the Average Participation scheme allocates subflows according to equation~\eqref{eq:transactions_to_flow_allocations}. As the underlying flow tracing departs from each source to trace the physical net flow to all sinks, and analogously for departure from each sink, it only accounts for flows in the prevailing net flow direction. Potential counter-flows that run opposite to that direction are not taken into account. 

Upstream and downstream tracing result in the same allocation if and only if no line losses are taken into account. In case of a nonzero line loss, downstream tracing considers gross flows going into a line, whereas the upstream setup only takes net flows coming out of a line. As the Average Participation scheme is fairly intuitive, it gained attention in the research community. Likewise, $AP$ tracing does not depend on physical laws specific to electricity networks. It thus readily applies also to other sorts of transport systems, for instance, in finance of logistics. Appendix \ref{sec:appendix_ap} details the mathematical setup leaning on the formulation by~\citet{achayuthakan_electricity_2010-1}. 


\subsubsection*{Marginal Participation}\label{sec:alloc_subsec:alloc_subsubsec:mp}
The Marginal Participation ($MP$) allocation scheme is based on a sensitivity calculation for the linear power flow formulation. It was first presented by~\citet{rudnick_marginal_1995}. 

The scheme measures how power flows vary when injection at node~$n$ is infinitesimally increased. This infinitesimal increase is absorbed by all other nodes. These act as slack node(s) proportional to their contribution to the total absolute net injection in the network. In this respect, the proportion how much each node contributes to the slack may vary across the analyzed time steps. The difference between the flow pattern based on the marginal increase at a node and the original flow pattern is the sensitivity value. This value is multiplied with the nodal power injection, resulting in virtual injection patterns~$\allocateInjection$ and, following equation~\eqref{eq:alloc_vip_to_allocflow}, the according partial flows~$\allocateFlow$. For the virtual injection patterns, we implicitly assume a fifty-fifty split between sources and sinks because all other nodes serve as sinks for the infinitesimal increase, irrespective of being a source or sink in the original flow pattern. Appendix~\ref{sec:appendix_mp} presents further mathematical details. 

The allocation scheme is based on the linear power flow formulation and assumes no power losses. The derivations of the peer-to-peer transactions and the according flow allocation is carried out separately for each node. Therefore, the Marginal Participation scheme may specify potential counter-flows. 


\subsubsection*{Equivalent Bilateral Exchanges}\label{sec:alloc_subsec:alloc_subsubsec:ebe}
The Equivalent Bilateral Exchanges ($EBE$) allocation assumes that every source serves every sink proportionally to its total net demand. Vice versa, every sink is served by every source proportionally to its net production. $EBE$ flow allocation was first proposed by~\citet{galiana_transmission_2003}.

To implement $EBE$ allocation, the scheme takes one source~$n$, sets the injection of all other sources to zero, and linearly scales down the net demand of all sinks in the network until it matches the net export of~$n$. Based on the resulting dispatch, the virtual injection patterns~$\allocateInjection$ for node~$n$ are specified, and the resulting power flow~$\allocateFlow$ is allocated according to equation~\eqref{eq:alloc_vip_to_allocflow}. Analogously for each sink~$n$, the scheme sets the demand of all other sinks in the network to zero and scales down all net production until it matches the net demand of~$n$. Again, based on the virtual injection patterns~$\allocateInjection$, the flow resulting from this dispatch is the allocated flow~$\allocateFlow$ for node~$n$, according to equation~\eqref{eq:alloc_vip_to_allocflow}. For each emerging virtual injection pattern, there is a degree of freedom which proportion is allocated to the respective source(s) and sinks(s). As a default for our illustration and evaluation below, we impose a fifty-fifty split. 

As net sources and sinks are treated separately, the Equivalent Bilateral Exchanges scheme may also allocate counter-flows against the prevailing net flow direction resulting from the overall dispatch. Appendix~\ref{sec:appendix_ebe} provides further mathematical details.


\subsubsection*{Linearized Z-bus Flow Allocation}\label{sec:alloc_subsec:alloc_subsubsec:zbus}
The Z-bus flow allocation (\textit{Z-bus}) scheme was introduced by~\citet{conejo_z-bus_2007}. It is based on the full AC power flow formulation and breaks down the mathematical contributions of each nodal current injection to the active power flow on each line. This is possible because the active power flow on a line is given as the sum of nodal current contributions. Each single contribution can be considered as the physical impact of a current injection at~$n$ to the AC power flow on line~$\ell$. In Appendix~\ref{sec:appendix_zbus}, we show how the non-linear formulation can be linearized for applications in linear power flow models. Note that the~\textit{Z-bus} scheme is used by the Power Divider method~\citep{chen_power_2016}, which makes a separate consideration of the latter redundant. In its default approach, the \textit{Z-bus} allocation scheme allocates flows to sources only. Yet all other divisions between sources and sinks are possible. In our illustration and evaluation, we weigh sources and sinks equally.


\section{Working example and illustration}\label{sec:application}

\subsection{Working example: model description and snapshots}\label{sec:application_subsec:example}
\begin{figure}[b!]
    \centering
    \includegraphics[width=.6\textwidth]{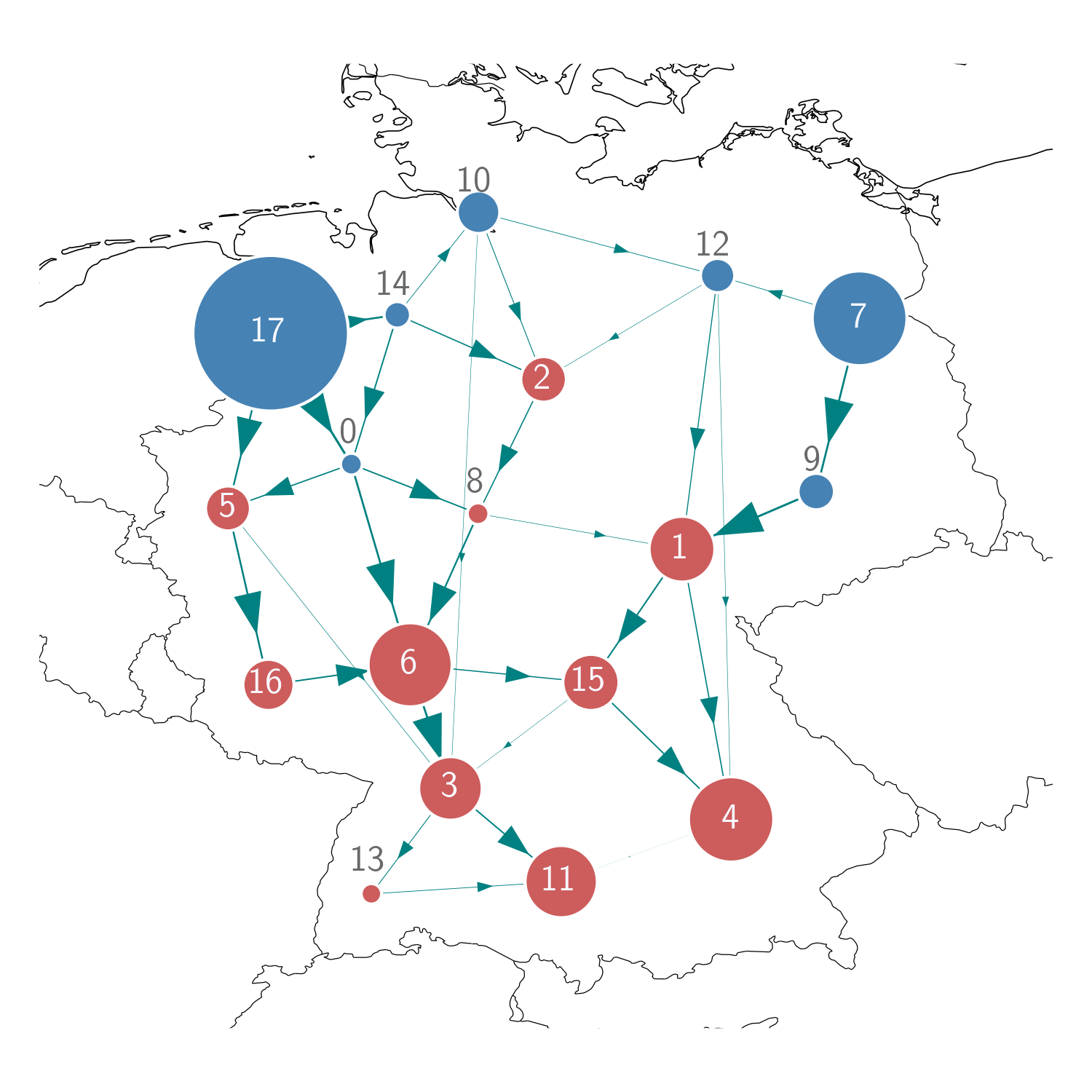}
    \caption{Working example: exemplary dispatch and flow within an aggregated German network model for 2030. Positive injections are blue, negative injections red, circle and arrow sizes are proportional to dispatch and flow. This snapshot, which we refer to as \textit{normalstate}, represents an average operational state of the system.}
    \label{fig:normalstate}
\end{figure}

Throughout this paper, we apply the schemes and illustrate the results using a working example for Germany in the year~2030. This working example does not claim to be a realistic forecast of the power system, but serves as a test case, loosely reflecting a plausible future German electricity system. To this end, we resort to an aggregated representation of the German power system. The network is derived from the open-source model PyPSA-EUR~\citep{horsch_pypsa-eur_2018} that includes all transmission lines and substations at and above a voltage level of~$220$\,V. We restrict the full model to nodes and lines that lie entirely in Germany and apply the K-means clustering algorithm to aggregate the system to~$18$ nodes and~$34$ lines. Offshore areas are added to the nearest onshore node. Figure~\ref{fig:app_topology} in Appendix~\ref{sec:appendix_network} provides a sketch of the network including node and line denotations.

Generation and transmission capacities for the study year 2030 are determined using the python package PyPSA~\citep{brown_pypsa:_2018}. As a linear program, it minimizes the cost of providing electricity for one year in hourly resolution. The model pursues a brownfield approach that takes into account existing capacities by the base year~2019 and determines optimal expansion and reduction of generation and transmission for the year~2030. The optimization takes into account the hourly dispatch and a range of techno-economic constraints as well as the German CO$_2$ target for the year~2030. \citet{pypsa-eur-doc}~provide a full documentation of the model and input data.  
A main characteristic of the network is a strong offshore wind power fleet, with an overall capacity of around~$49$\,GW connected to three northern nodes. Natural gas is the most important fossil supply technology ($46$\,GW). Of the total annual electricity demand of~$466$\,TWh, renewables supply about~$59$\,\%. Compared to the base year~2019, the transmission network is expanded by around~$40$\,GW, mostly in north-south direction.

Within the working example, for convenience, we resort to three snapshots, each based on one time step within the model. Figure~\ref{fig:normalstate} depicts an average operational state with a relatively even mix between renewable and fossil power production. Blue circles indicate a net injection, red circles a net withdrawal from the grid. The direction of arrows shows the flow direction on each line, their size indicates the flow volume. Additionally, the figure shows for each node its corresponding label. In the following, the shown snapshot will be referred to as~\textit{normalstate}.

\begin{figure}[t!]
    \centering
    \includegraphics[width=0.8\linewidth]{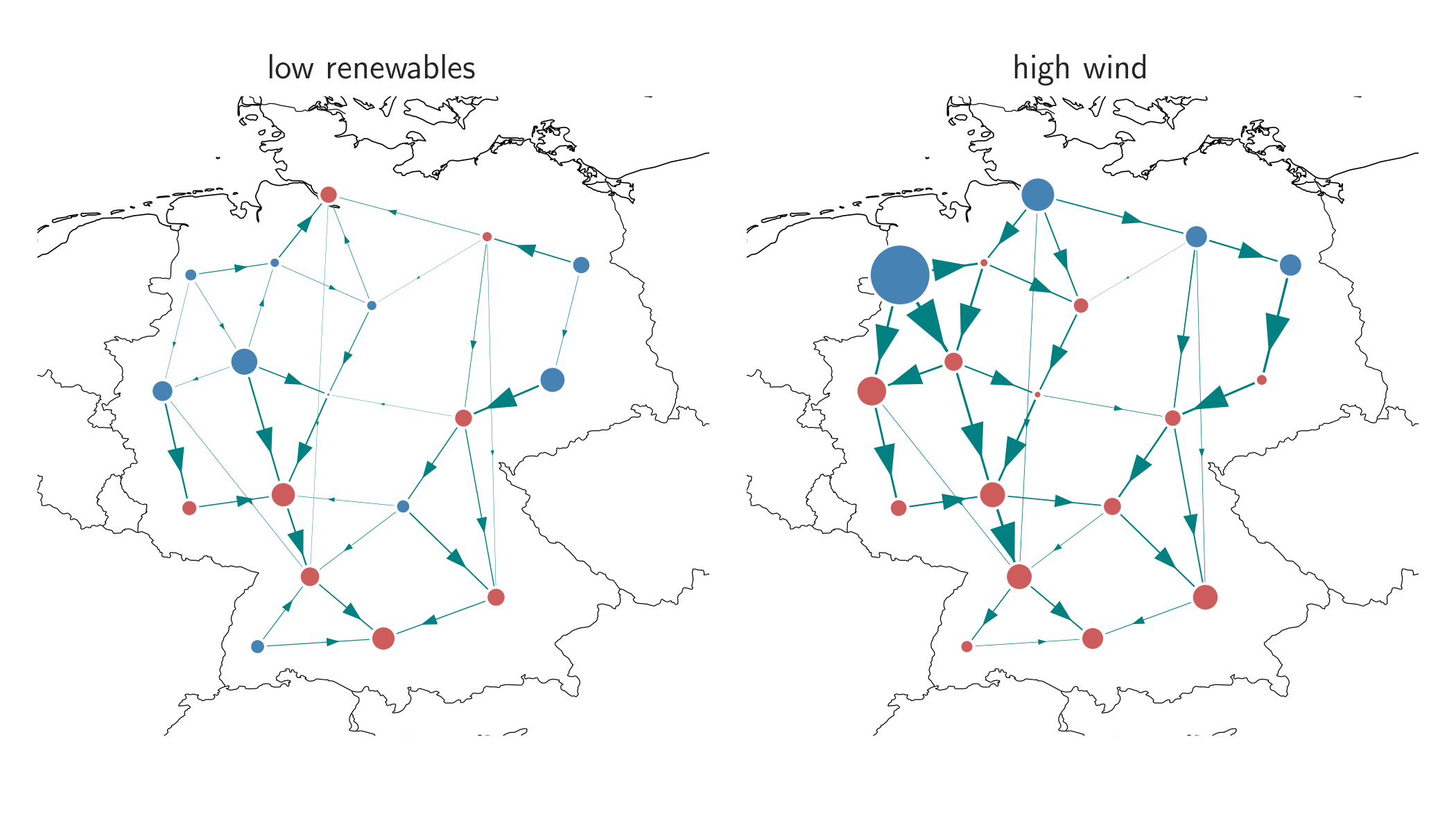}
    \caption{Working example: dispatch and flow within the aggregated German network in the snapshots \textit{low renewables} and \textit{high wind}. Positive injections are blue, negative injections red, circle and arrow sizes are proportional to dispatch and flow.}
    \label{fig:snapshots}
\end{figure}

Beside the \textit{normalstate}, we resort to two further snapshots that capture specific situations that occur in the power system. 
The snapshot~\textit{low renewables} is a fossil generation-based operational state with low renewable feed-in and high load; the snapshot~\textit{high wind} has over~$87$\,\% wind power in the system, mainly injected in the north of Germany. Figure~\ref{fig:snapshots} shows the respective dispatch and flow situations.

\subsection{Illustration of the allocation schemes}\label{sec:application_subsec:illustration}

\begin{figure}[t!]
\centering
\includegraphics[width=0.85\linewidth]{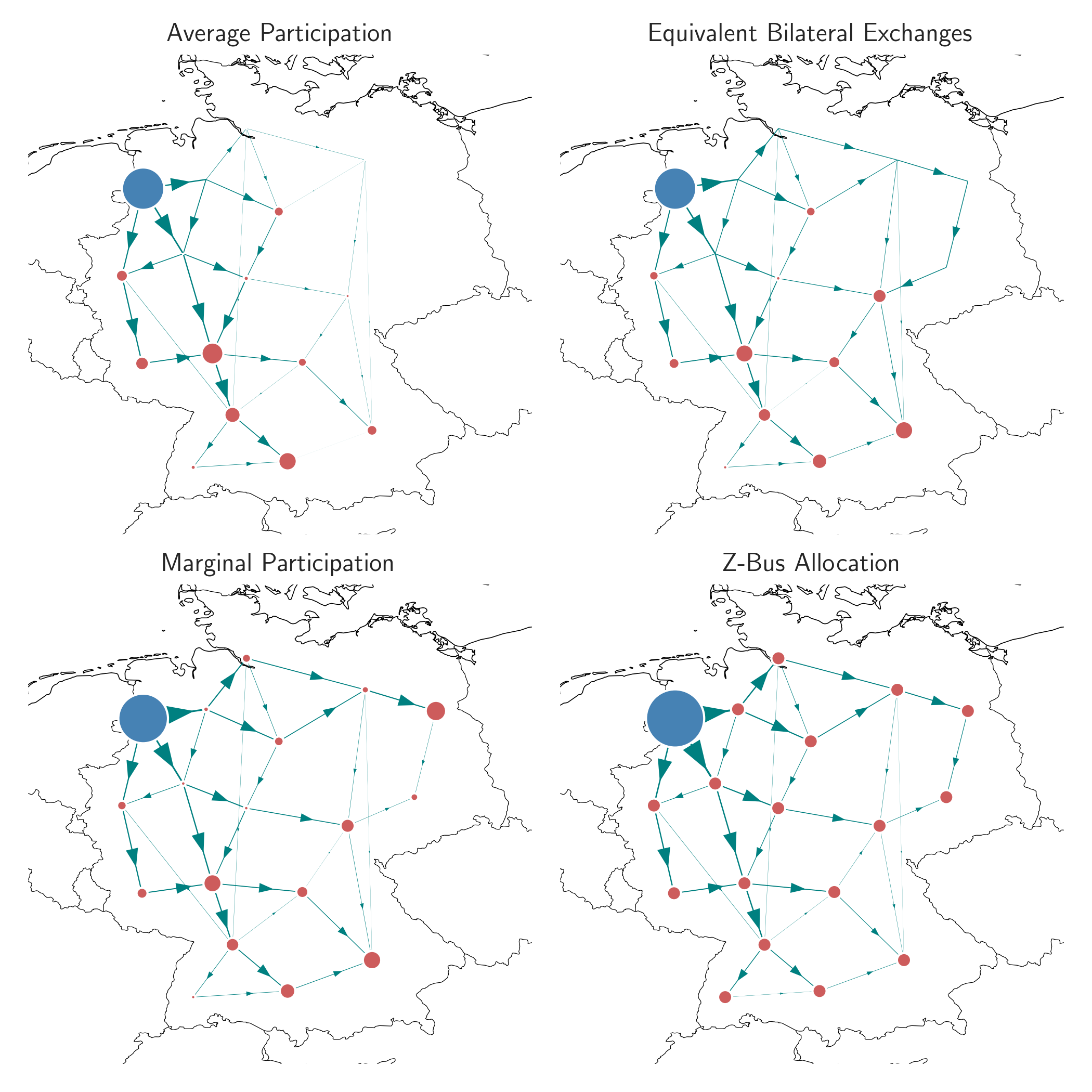}
\caption{Comparison of flow allocation schemes for the northwestern node~$17$ in the snapshot~\textit{normalstate}. The source is indicated in blue, sinks are drawn in red, circle and arrow sizes are proportional to net export/import and flow.}
\label{fig:compare_methods}
\end{figure}

The four schemes lead to significantly different allocations of flows to nodes. Figure~\ref{fig:compare_methods} shows for our working example how they allocate the export of the northwestern node~$17$ in the snapshot~\textit{normalstate}, in which the node has the highest net injection among all nodes ($\sim$10\,GW, compare Figure~\ref{fig:normalstate}). As above, exporting nodes are blue, sinks are red, the size of the circles is proportional to their net export/import, and the size of the arrows is proportional to the allocated flow on the respective line. 

The Average Participation allocation scheme results in the most locally constrained flows among all schemes. It restricts the allocated flows to lines in the west of Germany that direct power to the south. The outgoing flows from the northwestern node under consideration mainly meet net sinks that absorb incoming flows. At the same time, allocations opposite to the direction of the actual net flow, referred to as counter-flows, are not considered in the $AP$ allocation. Therefore, flows tend to appear more localizable than under the other schemes. 

Allocated flows according to the Equivalent Bilateral Exchange scheme range further over the network. As it allows counter-flows, sinks the east are taken into account to a greater extent. Yet, similar to $AP$ allocation, the $EBE$ scheme allows only actual sinks to absorb power in an allocation (compare Figure~\ref{fig:snapshots}). Nodes with net power injection, for instance the north-eastern node~$7$, may not be allocated as sinks. Therefore, the $EBE$ allocation also appears to be rather localizable.
The allocated flows in the Marginal Participation scheme allow both counter-flows and the allocation of flows to nodes that are net sources other than the originating node under consideration. This leads to more nodes in the east of Germany, which are actual net producers, to be specified as consumers of power of the left upper node. Graphical inspection again suggests that the accordingly allocated flows spread further in the network.

Eventually, the linearized Z-bus allocation scheme distributes the consumption of the power exports of the northwestern node~$17$ fairly equally across all nodes in the network. This leads to a rather broad flow allocation pattern. Note that this does not have to be necessarily the case for the common Z-bus allocation because the linearization neglects losses and reactive power flows.


\section{Evaluation criteria}\label{sec:eval}
Depending on the allocation scheme, each node gets allocated a different proportion of flows on all lines. This may have repercussions for the economics and policy of network planning. To this end, we develop the four techno-economic quantitative evaluation criteria \textit{(i) fairness}, \textit{(ii) plausibility}, \textit{(iii) uniqueness}, and \textit{(iv) stability}. In the following, we introduce the concepts, apply them to the working example, and discuss the results.


\subsection{Fairness}\label{sec:eval_fairness}


\subsubsection{Concept}\label{sec:eval_fairness_concept}
\begin{subequations}
In economics, a standard notion of fairness invokes the idea of adequacy, i.e., a balance of needs and use. We follow this adequacy notion and refer to a balance between the need of the network and the use of the network. The subject of fairness is a node. To operationalize it, we relate a metric of network use allocated to a node and a metric of network dependency of a node. 

Consider the allocated network use first. In principle, we can take on a gross or a net perspective. From a gross perspective, our metric relies on the sum of absolute gross flows allocated to a node~$n$ according to allocation scheme~$a$. To take into account the extent of overall nodal exchanges, we normalize the flows allocated to a node to the overall gross flow volume.
\begin{align}
\hat{\rho}_{n}^{a}\equiv\frac{\sum_{t}\left(\sum_{\ell}\left|\allocateFlow^{a}\right|\right)}{\sum_{t}\left(\sum_{m,\ell}\left|\allocateFlow[m]^a \right|\right)} \Forall n
\end{align}

Metric~$\hat{\rho}_{n}^{a} \in \left[0,1\right]$ renders the proportion of absolute gross network use of a node according to allocation scheme~$a$. Yet effective power flows on each line are netted in a meshed network. Either a node increases the net flow on a line or it decreases it. In the first case, the node contributes to effective net network use. In the latter case, the node reduces effective net network use. To account for the effects on net line flows, we specify a metric of allocated net network use or network stress: 
\begin{align}
&\rho^{a}_{n}\equiv \dfrac{\sum_{t}\left[\sum_\ell \text{sgn}\left(\flow \right)\,\allocateFlow^a\right]}{\sum_{t}\left(\sum_{\ell}\left|\flow\right|\right)} \Forall n
\label{eq:eval_fair_rho}
\end{align}

where~$sgn$ is the signum function that takes on~$1$ if its argument is positive, $-1$ if its argument is negative, and~$0$ if its argument is zero. The denominator of~\eqref{eq:eval_fair_rho} represents the total absolute net flows is the network. These may be alternatively expressed as the difference of allocated relieving and stressing flows. If the net flow on a line is positive, a node exerts stress on that line if its allocated flow is also positive, and it relieves stress if its allocated flow is negative -- and vice versa for negative net flows on that line. The numerator of~\eqref{eq:eval_fair_rho} specifies for each node its contribution to stressing and relieving flows according to allocation scheme~$a$. For instance, consider in a time step in which the net flow on line~$\ell$ is negative such that~$\text{sgn}\left(f_{\ell,t}\right)=-1$. If scheme~$a$ allocates a positive flow on that line to node~$n$, the node relieves network stress and metric~$\rho^{a}_{n}$ decreases. If scheme~$a$ allocates a negative flow on that line to node~$n$, the node contributes to network stress and metric~$\rho^{a}_{n}$ increases. Straightforward algebra shows that~$\sum_{n}\rho^{a}_{n}=1$. Thus, the entire network stress is broken down to nodes. If~$\rho^{a}_{n}>0$, then node~$n$ contributes to overall stress; if~$\rho^{a}_{n}>0$, then node~$n$ relieves overall stress because the allocated flows run counter the prevailing flow direction.

Metrics~$\hat{\rho}^{a}$ and~$\rho^{a}$ capture the allocated network use of nodes according to scheme~$a$. To invoke adequacy, we require a point of reference. A plausible choice is the network or trade dependency of a node. Some nodes are rather self-sufficient while others rely on exporting or importing electricity, depending on the balance of demand and supply that has historically grown or is assumed in future scenarios. From a nodal perspective, trade dependency does not take transit flows into account. To this end, we define nodal network dependency~$\tau_{n} \in \left[0,1\right)$ as the ratio of the absolute value of net injections~$\injection$ of a node and overall net injections in the network~\eqref{eq:eval_fair_trade}. In economic terms, we can interpret the injections as net power exports or net trade. Trade dependency results from the dispatch and is invariant to the allocation scheme.
\begin{align}
\tau_{n} \equiv \frac{\sum_{t}\left|\injection\right|}{\sum_{t,m}\left|p_{m,t}\right|} \Forall n \label{eq:eval_fair_trade}
\end{align}

Our fairness metrics~$\hat{\phi}_{n}^{a}$ and~$\phi_{n}^{a}$ relate the allocated network use or stress and the network dependency~\eqref{eq:eval_fair_metric}. They describe a balance between requirements and use of the grid that mirrors the notion of adequacy and fairness. If~$\phi^{a}_{n}=1$, then the proportion of network stress allocated to node~$n$ is equal to the proportion of network dependency. If~$\phi^{a}_{n}>1$, a node is allocated over-proportionate network stress; if~$\phi^{a}_{n} \in \left[\right.0,1\left.\right)$, it is allocated under-proportionate network stress. Finally, if~$\phi^{a}_{n}<0$, then a node mitigates network stress; the lower~$\phi^{a}_{n}$, the greater the extent of this mitigation in relation the node's network dependency.
\begin{align}
\hat{\phi}^{a}_{n} \equiv \frac{\hat{\rho}^{a}_{n}}{\tau_{n}}, \qquad \qquad \phi^{a}_{n} \equiv \frac{\rho^{a}_{n}}{\tau_{n}} \Forall n \label{eq:eval_fair_metric}
\end{align}

The metric informs analysts in at least two ways. First, it can hint toward potential imbalances in the network in a sense that certain nodes over-proportionately stress the network. This may indicate network expansion needs or may inform cost allocation schemes. Second, it can be aggregated over the entire network to assess whether an allocation scheme represents a balance of overall network needs and overall allocated network stress. To this end, we propose the root mean squared error with respect to deviations from an equitable network use reflected by a numerical value of one.
\begin{align}
\phi^{a}\equiv \sqrt{\frac{1}{n}\sum_{n}\left(\phi^{a}_n-1\right)^{2}}
\label{eq:global_fairness}
\end{align}

 
\subsubsection{Application to the working example}\label{sec:eval_fairness_example}

The left panel of Figure~\ref{fig:eval_fair_stress} shows the allocated net network use~$\rho^{a}_{n}$ for the snapshot~\textit{normalstate} for all nodes. Across allocation schemes, more than one third of net stressing flows is allocated to node~$17$ that connects offshore wind farms in the North Sea. About a further~$10$\% of flows are each allocated to nodes that collect demand centres in southern Germany ($4$, $11$) and the source Berlin ($7$). 

Also the other nodes with allocated flows around~$5$\% are located in the southern demand centers. The remaining nodes get allocated negligible flows. Only four nodes ($2$, $5$, $8$, $16$) also get allocated flows that mitigate network stress, yet only to a small extent. Thus, all allocation schemes allocate the bulk of stressing flows to nodes with either excess supply or excess demand. Across allocation schemes, the net network use is largely stable. While differences prevail, no scheme consistently allocates the largest or smallest net flows to single nodes. In this regard, schemes~$AP$ and~$EBE$ deliver rather similar results for many nodes. 

\begin{figure}[H]
    \centering 
    \includegraphics[width=.95\linewidth]{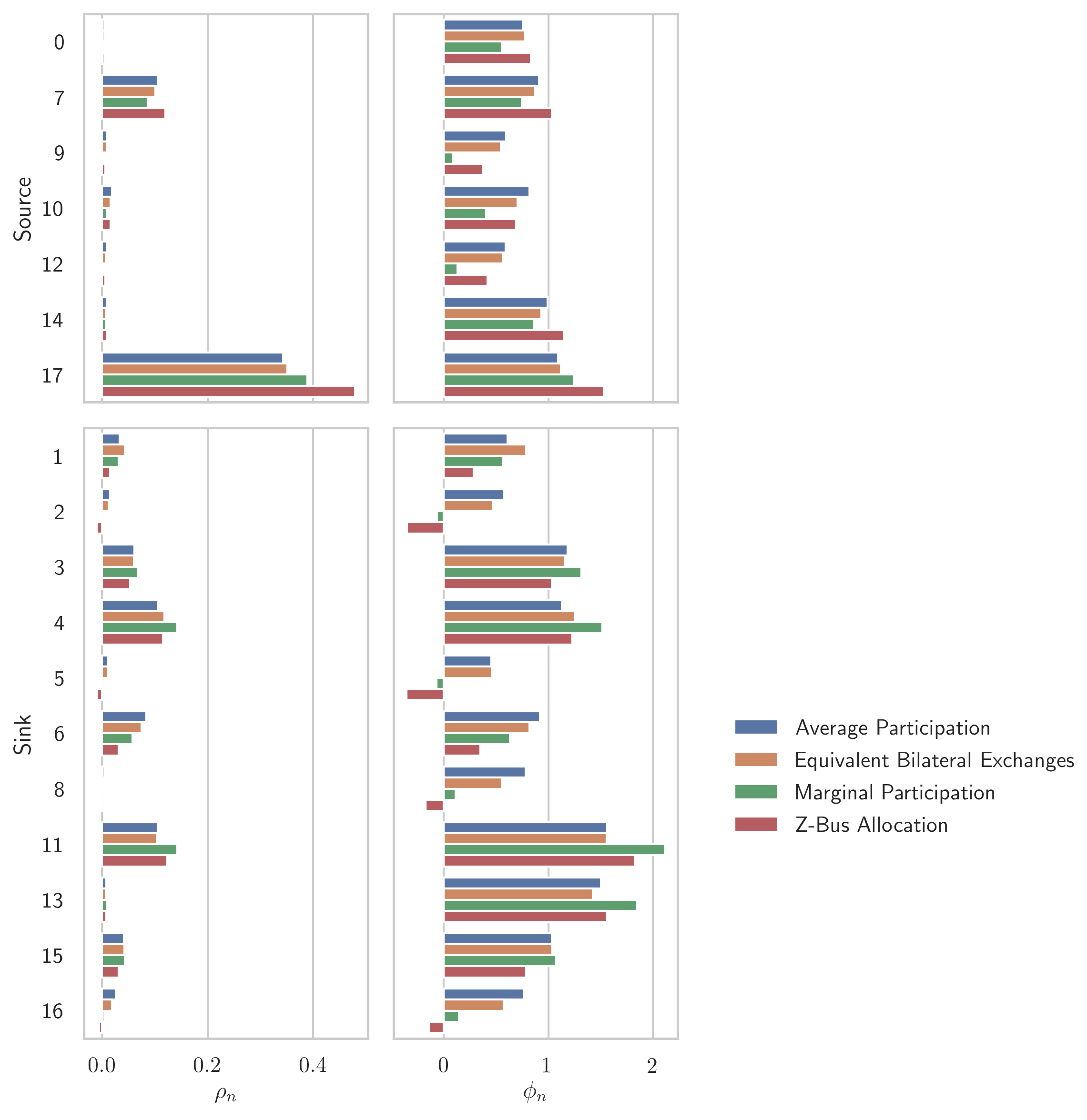}
    \caption{Allocated net flows~$\rho^{a}_{n}$ and fairness metric~$\phi^{a}_{n}$ for the snapshot~\textit{normalstate}}
    \label{fig:eval_fair_stress}
\end{figure}
To contrast allocated net network use with network needs, the right panel of Figure~\ref{fig:eval_fair_stress} shows the fairness metric~$\phi^{a}_{n}$ for all nodes for the snapshot~\textit{normalstate}. The offshore node~$17$ has a fairness metric between around~$1$ and~$1.5$, meaning that it gets allocated (slightly) more flows than its share in the absolute nodal exchange volume. The same applies to the southern demand centers at nodes~$3$ and~$4$. The southern demand center at node~$11$ gets allocated considerably more flows relative to its proportion in absolute nodal exchanges, and the eastern source~$7$ gets allocated less flows than its proportion in nodal exchanges. Node~$13$ sticks out as well. It gets attributed high network use relative to its share in nodal exchanges. While its allocated flows account for only around~$1$\% of net flows, its demand and supply-pattern is almost balanced.

\newpage
Comparing the metric across flow allocation schemes, two conclusions stick out. First, no flow allocation scheme allocates the highest or lowest values for the fairness metric consistently across nodes. Yet allocation schemes~$AP$ and~$EBE$ lead to comparable fairness metrics. Second, only the~$MP$ and linearized \textit{Z-bus} allocation schemes attribute a mitigating impact on network use for several nodes. By construction, they allow for the allocation of counter-flows; Average Participation only accounts for net positive flows. 

\end{subequations}

\subsection{Plausibility}\label{sec:eval_plaus}


\subsubsection{Concept}\label{sec:eval_plaus_concept}
\begin{subequations}
To evaluate plausibility, we refer to an intuitive understanding also for non-experts. In this spirit, a plausible allocation scheme~$a$ assigns to a node more flows on lines closer to it than farther away from it. Such plausibility measure may be node- or line-based. For a node-based definition, we relate the distance of all lines on which flows are allocated to a node. For a line-based definition, we relate the distance of all nodes that contribute to the allocated flow on a line. 

In both cases, operationalization requires a distance metric. A straightforward approach builds on the (geodesic) distance in a graph that renders the number of lines (edges) that connect two nodes (vertices) on the shortest path. As we require a distance metric between nodes and lines, we slightly adapt the geodesic distance. For a line, our metric renders a distance to a node equal to one if the line is directly connected to that node. The distance of a line to a node is equal to two if the line is not directly connected to the node in question but directly connected to a node that has a geodesic distance of one to the node in question, and analogously for all orders of distance~$k$. Similarly for a node, our metric renders a distance to a line equal to one if the line is directly connected to the node. The distance of a node to a line is equal to two if the node is not directly connected to the line in question but has a geodesic distance of one to a node that is directly connected to the line in question, and analogously for all orders of distance~$k$.

A distance matrix~$D$ of dimension~$(N \times L)$ helps to illustrate the network topology. Consider the sub-matrix~$\Tilde{D}$ capturing the first six nodes and lines in our working example~(compare Figure~\ref{fig:app_topology} for node and line denotations). For instance, the distance from node~$2$ to line~$1$, and vice versa the distance from line~$1$ to node~$2$, is~$2$ as represented by the element~$(3 \times 1)$.
\begin{align}
\Tilde{D} = \begin{bmatrix}
1 & 1 & 1 & 1 & 1 & 3 \\
3 & 3 & 3 & 3 & 2 & 1 \\
2 & 3 & 3 & 3 & 2 & 2 \\
3 & 3 & 2 & 2 & 3 & 3 \\
3 & 4 & 4 & 3 & 3 & 2 \\
2 & 2 & 1 & 2 & 2 & 4  
\end{bmatrix} 
\label{eq:eval:plaus_submatrix}
\end{align}

To operationalize the distance metric for our plausibility measure, we transfer the distance information to level-k distance vectors. For a node, the node distance vector~$d^{k}_n$ of dimension~$(L \times 1)$ has an entry of~$1$ if a line has a distance of~$k$ to that node and zero else. Analogously, the line distance vector~$d^{k}_l$ of dimension~$(N \times 1)$ has an entry of~$1$ if a node has a distance of~$k$ to that line and zero else. For our working example, the node distance (sub) vectors of node~$5$ for the first six lines are given by:

\begin{small}
\begin{minipage}[h]{.20\textwidth}
\begin{align}
\Tilde{d}^1_{5} = \begin{bmatrix}
0 \\
0 \\
1 \\
0 \\
0 \\
0 
\end{bmatrix}, \notag
\end{align}
\end{minipage}
\begin{minipage}[h]{.20\textwidth}
\begin{align}
\Tilde{d}^2_{5} = \begin{bmatrix}
1 \\
1 \\
0 \\
1 \\
1 \\
0 
\end{bmatrix}, \notag
\end{align}
\end{minipage}
\begin{minipage}[h]{.20\textwidth}
\begin{align}
\Tilde{d}^3_{5} = \begin{bmatrix}
0 \\
0 \\
0 \\
0 \\
0 \\
0 
\end{bmatrix}, \notag 
\end{align}
\end{minipage}
\begin{minipage}[h]{.25\textwidth}
\begin{align}
\Tilde{d}^4_{5} = \begin{bmatrix}
0 \\
0 \\
0 \\
0 \\
0 \\
1 
\end{bmatrix}  
\end{align}
\end{minipage}
\end{small}
\\

That is, node~$5$ has a distance of 1 to line~$3$, a distance of~$2$ to lines~$1$, $2$, $4$, and~$5$, a distance of~$3$ to no line, and a distance of~$4$ to line~$6$. Line-based distance vectors~$d^{k}_{\ell}$ are constructed in a analogous fashion. 

With the distance vectors at hand, we can relate the flow on all lines of distance~$k$ to each node~$n$ and divide it by the absolute flow allocated to that node. This constitutes the node-based plausibility metric~$\delta^{k,a}_{n}$ for allocation scheme~$a$~\eqref{eq:eval_plaus_phi_l}. Analogously, we relate the flow allocated to node of distance~$k$ to each line and divide it by the absolute flow on that line, rendering the line-based plausibility metric~$\delta^{k,a}_{\ell}$ for allocation scheme~$a$~\eqref{eq:eval_plaus_phi_l}.
\begin{align}
\delta^{k,a}_{n} \equiv \frac{\sum_{t}\left(\sum_{\ell}d^{k}_{n}\left|F^{a}_{n,\ell,t}\right|\right)}{\sum_{t}\left(\sum_{\ell}\left|F^{a}_{n,\ell,t}\right|\right)} \label{eq:eval_plaus_phi_n} \Forall k,n \\
\delta^{k,a}_{\ell} \equiv \frac{\sum_{t}\left(\sum_{n}d^{\ell}_{k}\left|F^{a}_{n,\ell,t}\right|\right)}{\sum_{t}\left(\sum_{n}\left|F^{a}_{\ell,t}\right|\right)} \Forall k,\ell \label{eq:eval_plaus_phi_l}
\end{align}

Metrics~$\delta^{k,a}_{n}$ and~$\delta^{k,a}_{\ell}$ are specific to each node and line, respectively. For a statement on the entire network, we can aggregate over all nodes or lines, weighted by their proportion of total allocated flows in the system. By construction, the resulting metric~$\delta^{k,a} \in \left[0,1\right]$ is identical.
\begin{align}
\delta^{k,a} \equiv \frac{\sum_{n,\ell,t}d^{k}_{n}\left|F^{a}_{n,\ell,t}\right|}{\sum_{n,\ell,t}\left|F^{a}_{n,\ell,t}\right|} 
= 
\frac{\sum_{n}\frac{\sum_{\ell,t}d^{k}_{n}\left|F^{a}_{n,\ell,t}\right|}{\sum_{\ell,t}\left|F^{a}_{n,\ell,t}\right|}\sum_{\ell,t}\left|F^{a}_{n,\ell,t}\right|}{\sum_{n,\ell,t}\left|F^{a}_{n,\ell,t}\right|}
=
\frac{\sum_{n}\left(\delta^{k,a}_{n}\sum_{\ell,t}\left|F^{a}_{n,\ell,t}\right|\right)}{\sum_{n,\ell,t}\left|F^{a}_{n,\ell,t}\right|} \Forall k \\
\delta^{k,a} \equiv \frac{\sum_{n,\ell,t}d^{k}_{\ell}\left|F^{a}_{n,\ell,t}\right|}{\sum_{n,\ell,t}\left|F^{a}_{n,\ell,t}\right|} 
= 
\frac{\sum_{\ell}\frac{\sum_{n,t}d^{k}_{\ell}\left|F^{a}_{n,\ell,t}\right|}{\sum_{n,t}\left|F^{a}_{n,\ell,t}\right|}\sum_{n,t}\left|F^{a}_{n,\ell,t}\right|}{\sum_{n,\ell,t}\left|F^{a}_{n,\ell,t}\right|}
=
\frac{\sum_{\ell}\left(\delta^{k,a}_{\ell}\sum_{n,t}\left|F^{a}_{n,\ell,t}\right|\right)}{\sum_{n,\ell,t}\left|F^{a}_{n,\ell,t}\right|} \Forall k
\end{align}

In general, it holds that~$\sum_{k}\delta^{k,a}=1$ because each flow is allocated to some level of distance. For instance, consider for an allocation scheme the result~$\delta^{1,a}=0.3$ and~$\delta^{2,a}=0.7$. Then, $30$ percent of all flows occurring in the network are allocated to nodes (lines) of level~$1$, that is, directly connected to the lines (nodes), and~$70$ percent of all flows are allocated to nodes (lines) of level~$2$. The more flows are allocated to nodes in closer distance, the more localized the result of a flow allocation scheme and the more intuitive or plausible it may be perceived.

Finally, aggregating over distance levels renders one dense metric for each allocation scheme~$a$. This may be specific to a node or aggregated over the entire network. To this end, we define~$\delta^{a}_{n}~\in~\left[1,k\right]$ and~$\delta^{a}~\in~\left[1,k\right]$ as the weighted average distances of all allocated flows.
\begin{align}
\delta^{a}_{n} \equiv \sum_{k}k\delta^{k,a}_{n} \Forall n, \qquad \delta^{a} \equiv \sum_{k}k\delta^{k,a}
\label{eq:global_plausibility}
\end{align}


\subsubsection{Application to the working example}\label{sec:eval_plaus_example}

Figure~\ref{fig:eval_plaus_overall} shows the average distance of allocated flows for the four allocation schemes in our working example. For each snapshot, average participation allocates the shortest distance between flows and nodes, and~$EBE$ allocates the greatest distance. Overall, the distances range between~$1.6$ and~$2.4$. Concerning the snapshots, allocated distances are shortest for snapshot~\textit{low renewables} and greatest for snapshot~\textit{normalstate}. While allocated distances vary relatively strongly between snapshots for schemes~$AP$ and~$EBE$, they are quite constant for the linearized \textit{Z-bus} scheme.

Upon closer inspection, the average distances both informs about characteristics of the network as well as the respective snapshots and reflect the characteristics of the allocation schemes. For the Average participation allocation scheme, average flow distances are, generally, short. Due to construction, it traces flows from each sink (source) through the network that get (partially) absorbed when they meet a source (sink). Additionally, there are no counter-flows that may maintain an allocated gross flow on a line in one direction although the net flow runs in the opposite direction. Therefore, on average, allocated flows tend to be closer to the node in question than for the other schemes. This is also in line with the graphical intuition developed in Figure~\ref{fig:compare_methods}. 

\begin{figure}[b!]
\centering 
\includegraphics[width=.95\linewidth]{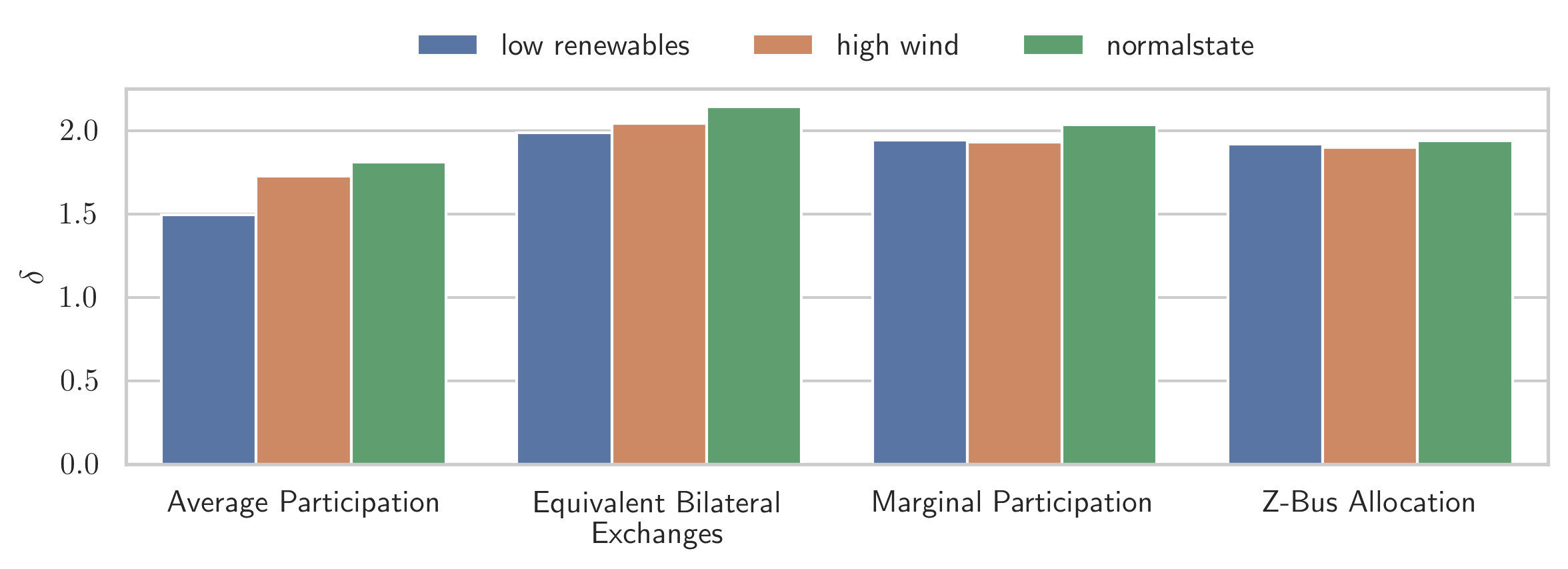}
\caption{Plausibility metric~$\delta^{a}$ that indicates the average distance of allocated flows for the four allocation schemes~$a$ under consideration}
\label{fig:eval_plaus_overall}
\end{figure}

Results for the Average Participation scheme also reflect the characteristics of the network and snapshots. For~$AP$ allocation, the average distance is lowest for the \textit{low renewables} snapshot and highest for the \textit{normalstate} snapshot. In the \textit{normalstate}, there is high wind supply in the north and flows mainly head to load centers in the west and south. This prevailing flow direction leads to higher average distances also for~$AP$ allocation. As shown in \cite{hofmann_principal_2018} for snapshots with substantial renewable feed-ins, electricity flows tend to be aligned across wide areas. For the $AP$ scheme, these aligned flows will allow far reaching peer-to-peer relations, which, on average, push the distance. In contrast, sinks and especially sources are more evenly distributed for \textit{low renewables}. As there is no dominant flow direction, average distances of allocated flows are short, and in particular short for the $AP$ scheme. 

For the $MP$ allocation scheme, average distances are higher than for $AP$ allocation because it allows for allocated counter-flows. Compared to the $AP$ scheme, this particularly increases the average distance for the \textit{low renewables} snapshot. In the other two snapshots, average distances are also longer compared to $AP$ allocation yet to a lesser extent. This is so because the scheme considers sources as sinks when specifying peer-to-peer relations for a node. In the snapshots \textit{normalstate} and \textit{high wind}, northern nodes with large wind supply are partly allocated as sinks for each other, mitigating the increase in average distances.

For the~$EBE$ allocation scheme, average flow distances are, generally, long. By construction, it specifies peer-to-peer relations between sources and sinks based on their relative weight in overall nodal exchange. In the working example, dominant renewable sources are located at the northern nodes where a lot of wind power is connected, and dominant sinks are in the west and south. Especially for the snapshots \textit{high wind} and \textit{normalstate}, the allocation resulting from the $EBE$ scheme reflects these long distances (see also Figure~\ref{fig:compare_methods}) compared to allocation schemes $AP$ and $MP$. In contrast, the average distance for \textit{low renewables} is not much longer than under $MP$ and linearized \textit{Z-bus} allocation because in this snapshot dominant sources and sinks are geographically less dispersed. In general, also for other networks, longer distances are more likely than under $AP$ allocation because $EBE$ allocation takes all nodes in the network into account when specifying peer-to-peer relations.

For the linearized \textit{Z-bus} allocation scheme, average distances are both similar across snapshots and relatively low in comparison to the other schemes. We conjecture that this result is an artifact of the network in the working example, in which the longest distance from a node to a line is~$5$. By construction, linearized \textit{Z-bus} allocation should tend to specify longer distances because sinks are taken into account more evenly across the network due to an assumed uniform voltage level when linearizing the \textit{Z-bus} approach.

\end{subequations}

\subsection{Uniqueness}\label{sec:eval_unique}
Uniqueness of a flow allocation can be understood in two ways. First, as a binary indicator, it specifies whether an allocation scheme generates the same flow allocation when applied two times on the same set of spatial dispatch decisions. This is the case for all four schemes under consideration as there is no inherent ambiguity when tracing flows, specifying peer-to-peer relations or specifying virtual injection patterns.

Yet, second, there is a relevant degree of freedom when splitting the contribution between the supplying and demanding node in a peer-to-peer relation, compare equation~\eqref{eq:transactions_to_flow_allocations}, or virtual injection pattern. This applies to schemes Average Participation, Marginal Participation as well as Equivalent Bilateral Exchanges. While we present a fifty-fifty split as focal default, all other splits are possible upon researcher discretion. Also the linearized \textit{Z-bus} scheme implicitly assumes an equal split (compare Appendix~\ref{sec:appendix_zbus}).

This becomes especially relevant when basing economic decisions or policy measures on a flow allocation, or according cost allocation. Concerning the allocation of network costs, the flow split is related to the concept of G- and L-components that specify which fractions of network costs must be paid by generation (G) and load (L). As such, the three allocation schemes $AP$, $MP$, and $EBE$ can accommodate the full range between generation and load. In any case, for each allocation result, the underlying assumption must be communicated.  


\subsection{Stability}\label{sec:eval_stability}


\subsubsection{Concept}\label{sec:eval_stability_concept}
\begin{subequations}
We want to analyze whether and how allocated flows change as a response to a change in the underlying physical flows. The lower such change, the more stable results an allocation schemes delivers. Given a spatial dispatch pattern, we increase the net injection at one node~$m$ by increment~$i$ and decrease the net injection by the same increment at another node~$mm \neq m$. We then recalculate the resulting physical flows and re-allocate flows to nodes according to scheme~$a$. The resulting difference~\eqref{eq:eval_stable_1} is the basis for systematic stability comparisons.
\begin{align}
\Delta^{a}_{n,\ell,t}(i,m,mm) \equiv F^{a}_{n,\ell,t}(i,m,mm) - F^{a}_{n,\ell,t} \Forall n,\ell,t \label{eq:eval_stable_1}
\end{align}

Metric~$\Delta^{a}_{n,\ell,h}(i,m,mm)$ informs about the reaction of the flow allocation to changes in the net injection pattern of specific nodes. To make statements about the reaction to changes in the entire network, we average over all permutations of net injection changes. The absolute value ensures that stability refers to absolute changes; otherwise, positive and negative changes may net out.
\begin{align}
\Delta^{a}_{n,\ell,t}(i) \equiv \frac{\sum_{m,mm} \left| \Delta^{a}_{n,\ell,t}(i,m,mm)\right|}{N(N-1)} , \quad m \neq mm \Forall n,\ell,t, 
\end{align}

where~$N$ is the number of nodes in the network. Summing over all nodes, lines, and time steps aggregates information for the entire network in an observation period. As absolute changes are unhandy to interpret, we relate this sum to the sum of absolute flows on all lines absent any perturbation.
\begin{align}
\Delta^{a}(i) \equiv \frac{\sum_{n,\ell,t}\Delta^{a}_{n,\ell,t}(i)}{\sum_{n,\ell,t}\left|F^{a}_{n,\ell,t}\right|}
\label{eq:global_stability}
\end{align}

Metric~$\Delta^{a}(i)$ refers to the overall stability of an allocation scheme~$a$. It specifies the average normalized absolute deviation from the initial flow allocation for increment~$i$. The larger~$\Delta^{a}(i)$, the more do allocated flows react to a perturbation of the original pattern of network injections and withdrawals. As such, a small increment of, say, $1$~MW appears plausible in many applied settings to detect whether a small change in actual flows results in a large change in allocated flows. Varying the increment offers another approach to analyze whether an allocation is stable. As~$\Delta^{a}(i)$ is a function of the increment~$i$, a convex curve, for instance, would indicate rather high stability of the allocation for small changes in actual flows and rather low stability for large changes.


\subsubsection{Application to the working example}\label{sec:eval_stability_example}
Figure~\ref{fig:eval_stable} shows the average deviation~$\Delta^{a}(i)$ of absolute allocated flows in percent in response to perturbations of~$i=1$\,MW, $i=10$\,MW, and~$i=100$\,MW for the three snapshots and four allocation schemes. For a~$1$\,MW perturbation, the deviation is below~$0.01$\% in any case; for a~$10$\,MW perturbation, the deviation is below~$0.1$\%; and for a~$100$\,MW perturbation, the deviation is below~$1$\%. Comparing allocation schemes~$AP$, $MP$, and~$EBE$, the stability metric is fairly similar within each snapshot and perturbation. Likewise, the increase in response to a perturbation is almost perfectly linear. For a tenfold perturbation, the stability metric is about ten times higher. Thus, none of the three allocation schemes performs notably different with respect to stability and is clearly preferable concerning its overall stability to changes in underlying flows. In contrast, the linearized \textit{Z-bus} allocation scheme sees a higher stability for larger perturbations. While metric~$\Delta^{zbus}$ is around~$0.01$\,\% for a perturbation of~$1$\,MW, as for the other three allocations schemes, it is around half for larger perturbations.
\begin{figure}[t!]
    \centering 
    \includegraphics[width=\linewidth]{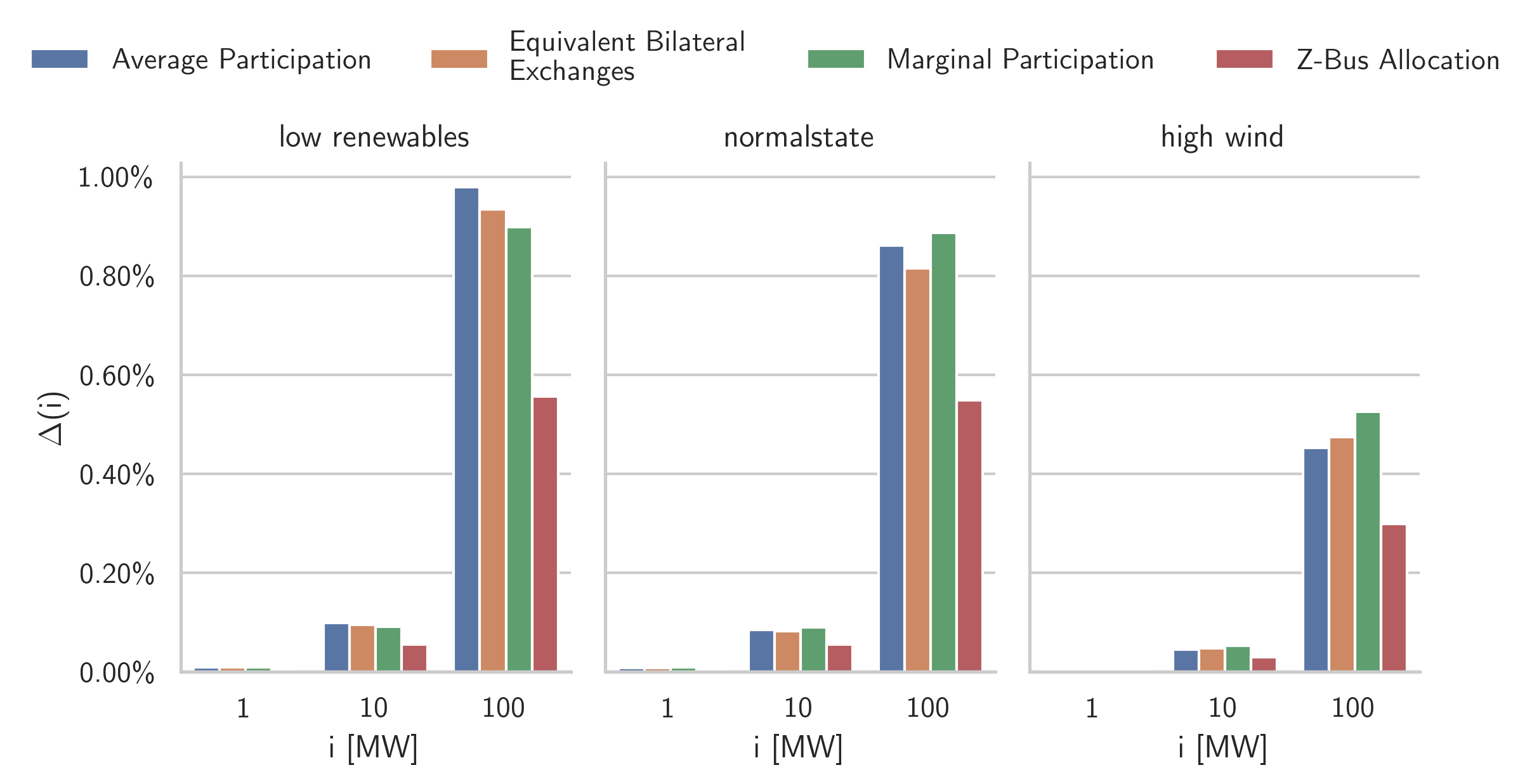}
    \caption{Stability metric~$\Delta^{a}$ of the allocation schemes for perturbations~$i$ of~$1$\,MW, $10$\,MW, and~$100$\,MW for the three snapshots}
    \label{fig:eval_stable}
\end{figure} 

\end{subequations}


\section{Robustness: evaluation criteria and nonlinear AC load flow}\label{sec:eval_acdc}
We applied our quantitative evaluation criteria to the power flows resulting from a linear DC load flow approximation, as widely used in applied research. Yet the quantitative evaluation results may be sensitive to this simplification. Therefore, we additionally apply our evaluation criteria to a full nonlinear AC power flow setting for the snapshot \textit{normalstate}. To this end, we take the dispatch and demand pattern as initial condition for calculating the AC power flow using the Newton-Raphson algorithm. The resulting nonlinear power flow differs from the linear case within a relative margin of~$5$\%. In absolute terms, the largest differences occur in the north-west of the system. 

\begin{figure}[t!]
\centering
\includegraphics[width=0.75\textwidth]{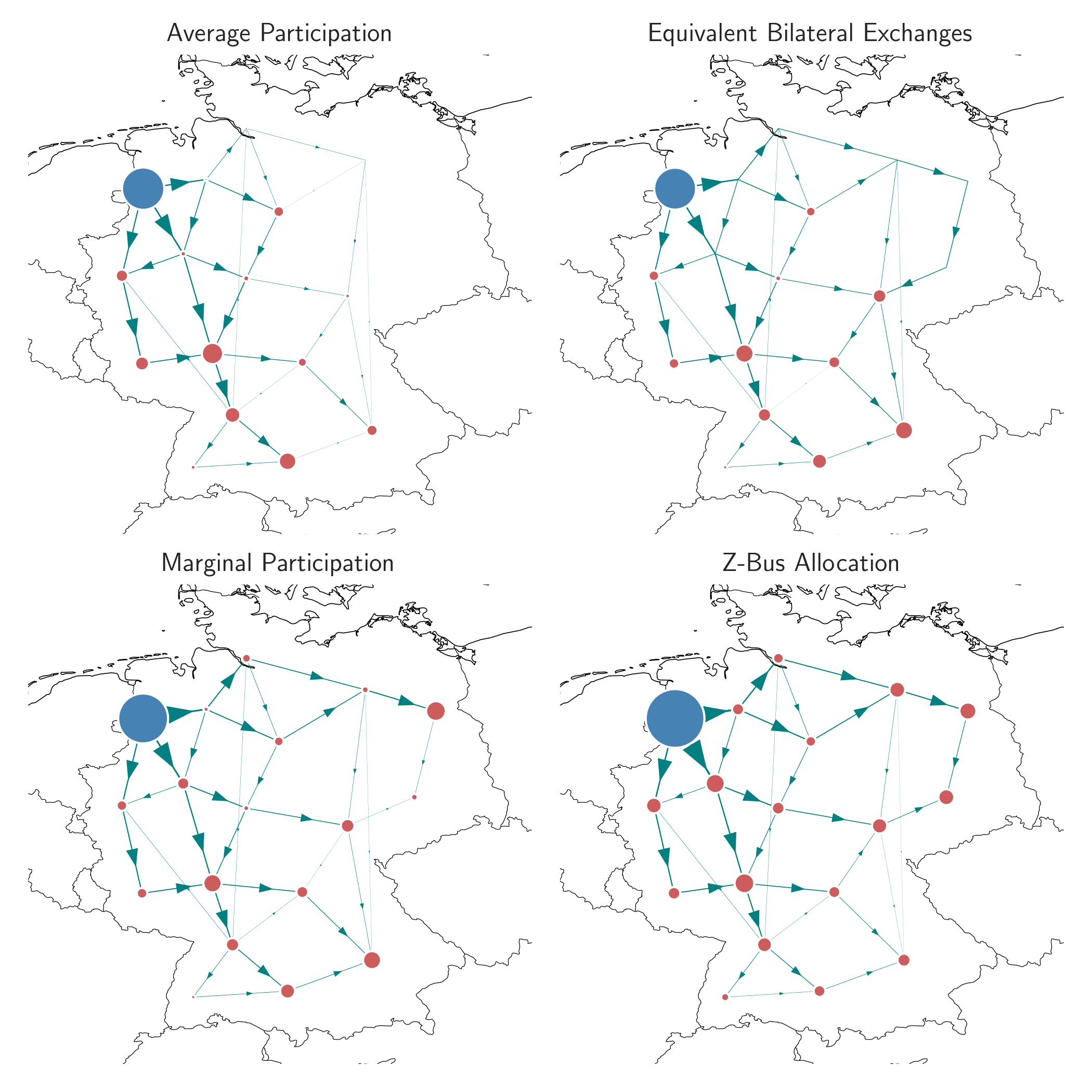}
\caption{Comparison of the flow allocation schemes based on the full nonlinear AC power flow for the northwestern node~$17$ in the snapshot \textit{normalstate}. The source is indicated in blue, sinks are drawn in red, circle and arrow sizes are proportional to net export/import and flow.}
\label{fig:compare_methods-nonlinear}
\end{figure}

Analogous to Figure~\ref{fig:compare_methods} for the linear approximation case, Figure~\ref{fig:compare_methods-nonlinear} shows the allocated flows according to the four schemes under consideration for northwestern node~$17$ for the snapshot \textit{normalstate} for the full nonlinear AC power flow calculation. Upon visual inspection, the schemes \textit{AP} and \textit{EBE} do not suggest noteworthy differences from the linear case. In contrast, the flow allocation based on the \textit{Z-bus} scheme exhibits noticeable differences to the linear case. While the scheme allocates the injected power of node~$17$ evenly among all nodes for the linear load flow approximation, it rather takes buses in closer vicinity into account for the nonlinear AC load flow case. Further, a high share of flows is allocated to other sources with strong injections such as, for instance, node~$7$ in the east. Overall, for the AC load flow case, the \textit{Z-Bus} scheme shows results similar to the \textit{MP} scheme, but allocates less flows between sources and strong net consumers. 

\begin{figure}[t!]
\includegraphics[width=0.8\textwidth]{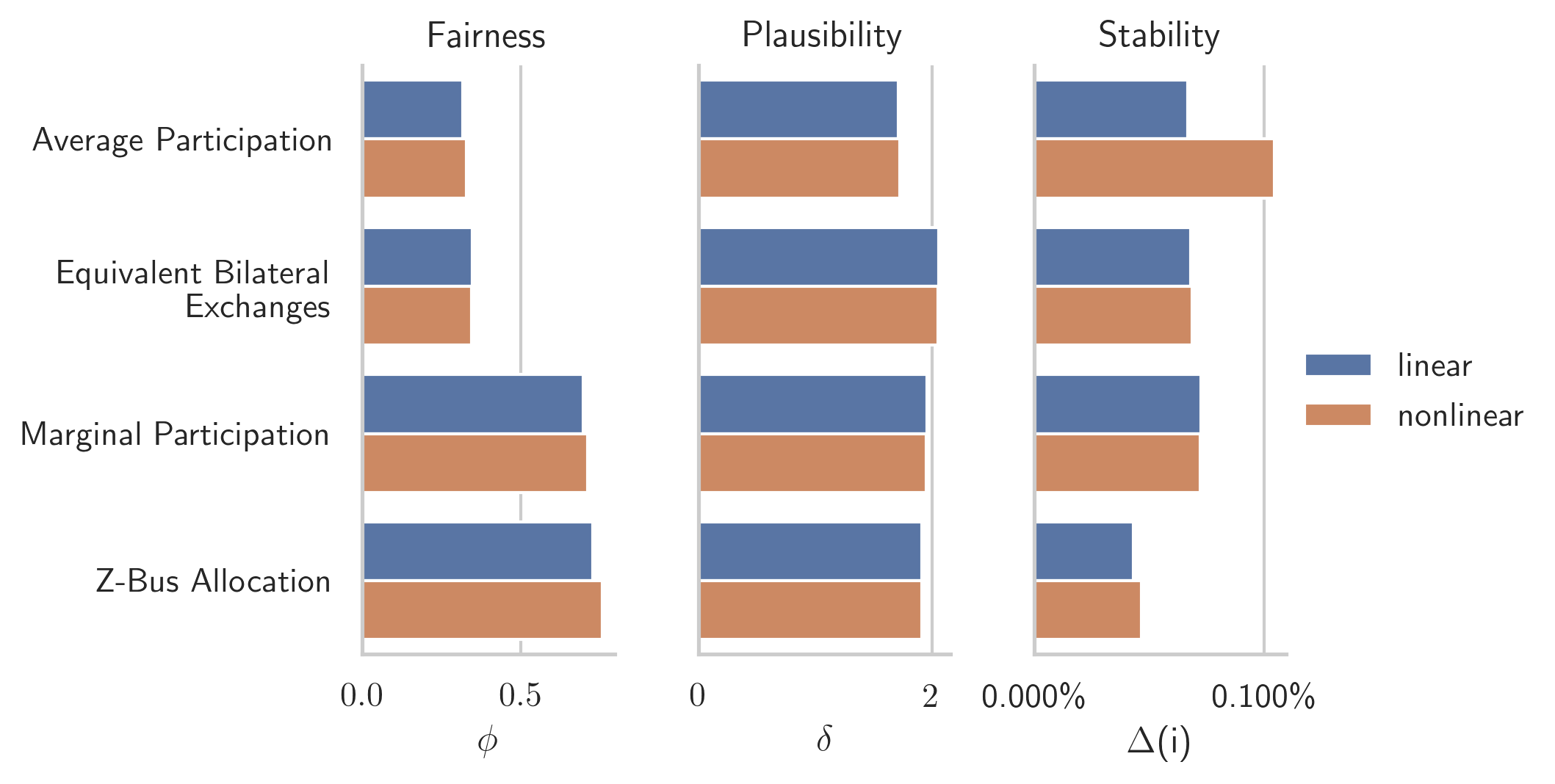}
\caption{Comparison of the fairness metric~$\phi$, the plausibility metric~$\delta$, and the stability metric~$\Delta(i)$ for the linear DC load flow approximation and the nonlinear full AC load flow calculation (snapshot \textit{normalstate}). For the stability analysis, the increment is set to $i$~=~10~MW.}
\label{fig:global_comparison}
\end{figure}

\newpage As the allocated flows based on a full nonlinear AC calculation differ from the linear DC approximation, also the quantitative results of the evaluation criteria may differ to some extent. To this end, Figure~\ref{fig:global_comparison} compares the three quantitative metrics \textit{fairness}, \textit{plausibility}, and \textit{stability} for the snapshot \textit{normalstate}. 

For both the linear DC approximation and nonlinear AC calculation, the overall fairness metric~$\phi^a$, calculated using equation~\eqref{eq:global_fairness}, leads to similar results. Allocation schemes \textit{AP} and \textit{EBE} weigh network use against network needs in a comparably equitable fashion. Differences between the linear DC load flow approximation and the nonlinear AC power flow are small. In contrast, \textit{MP} and \textit{Z-Bus} indicate a more imbalanced allocation of network use and network need, slightly amplified for the nonlinear case.

Similarly for the plausibility metric~$\delta^{a}$, calculated according to equation~\eqref{eq:global_plausibility}, that captures the average distance of allocated flows, the differences between the DC approximation and the nonlinear AC flow calculation are marginal. This is also the case for the \textit{Z-bus} allocation scheme although the visual inspection may suggest otherwise.

The largest difference prevails for the stability metric~$\Delta(i)$ for the \textit{AP} allocation scheme. A perturbation of~$i=10\,$MW, iterated through all nodes, yields a deviation of allocated flows of around~$0.05$\%, when based on the linear load flow approximation, and to a deviation of around~$0.1$\%, when based on the nonlinear full AC flow calculation. For the other allocation schemes, the differences are smaller. Specifically, \textit{AP} flow allocation is based on tracing flows through the network and takes into account only the net flow direction on a line. Using the full nonlinear AC flow calculation, the flow direction on some lines may change. Accordingly, the stability metric is prone to larger changes. These can be amplified by the algorithmic Newton-Raphson approach to determining the AC flow pattern because it cannot determine the exact flows and may thus yield greater deviations from the arithmetically ``exact'' solution of the linear load flow approximation.

\section{Conclusion}\label{sec:conclusion}

The electricity transmission network has a central mediating role between supply and demand. In AC networks, flows cannot be directed but spread according to physical laws. There is no unanimous solution that determines which node is responsible for flows on a line and to which extent. Moreover, even if a flow on a line is allocated to nodes, a further degree of freedom exists how to weigh the contribution of its source and sink. Therefore, dedicated schemes are required to allocate flows to nodes and to specify who should be deemed responsible for what part of a flow on a line. Yet a systematic evaluation of allocation schemes according to quantitative techno-economic metrics is missing so far.

In this paper, we develop quantitative evaluation criteria concerning $(i)$ fairness, $(ii)$ plausibility, $(iii)$ uniqueness, and $(iv)$ stability. We apply them to four relevant flow allocation schemes from the literature -- Average Participation, Marginal Participation, Equivalent Bilateral Exchanges, and Linearized Z-bus Allocation -- for a mid-term future scenario for Germany as well as illustrate and discuss results. For transparency, reproducibility, and as a service, we provide an open-source package for implementing the four discussed flow allocation schemes in a numerical power sector model under a permissive license.

In the first place, the resulting power flow allocations and their evaluation inform about both the role of single nodes and lines in the network and the characteristics of the network topology as well as the temporal snapshots under consideration. In the second place, power flow allocation schemes may provide relevant information for the design of network policies or economics regulation, such as the allocation of costs for network maintenance or expansion. These may be based on allocated network use, potentially also in relation to network needs. We leave this latter aspects open for future research.

Based on our discussion, illustration, and evaluation, we refrain from recommending to use any specific flow allocation scheme in applied research or policy studies. Yet we highlight that there are differences and how they arise. In this respect, central conclusions are, first, that flow allocation patterns are, as such, unique for each discussed scheme, but entail a degree of (researcher) freedom how to distribute flows between supply and demand for each tuple of nodes. When confronted with flow allocation results, or farther-reaching cost allocation results based on them, researchers and analysts should be aware of the implicit assumption on the weighting of the supply and demand sides responsible for each allocated flow. Second, especially the Average Participation scheme tends to allocate more localizable flows, that is, flows on a line tend to be allocated to nodes closer to it. In contrast, the Equivalent Bilateral Exchanges allocation scheme tends to allocate flows over greater distances in our application as it is mainly based on the relative weight of sources and sinks among the total nodal exchange volume. In our working example, net sources are more in the north of the country, net sins in the south. When confronted with a flow allocation and its evaluation, a researcher or analyst must thus relate it to the characteristics of the network in question, specifically the location of supply and demand centers as well as potential prevailing flow directions. Third, all analyzed allocation schemes are stable with respect to systematic perturbations. The volume of changes in nodal injections translate almost linearly to the volume of changes in allocated flows, and even under-proportionately for the linearized \textit{Z-bus} allocation scheme. Thus, we see no clear drawback of using any scheme concerning its stability. 


\vfill
\noindent 
\textbf{Acknowledgments:} 
We thank Alexander Kies, Matthias Hanauske, and Markus Schlott for valuable comments. We gratefully acknowledge financial support within the project Net-Allok, funded by the Federal Ministry of Economic Affairs and Energy (BMWi) (grant number 03ET4046A).
\\

\noindent 
\textbf{Declarations of interest:} none.
\\

\noindent
\textbf{Role of the funding source:} The funding source had no involvement in study design, collection, analysis and interpretation of data, in the writing of the report, and in the decision to submit the article.

\newpage


\section*{References}
\printbibliography[heading=none]


\clearpage
\appendix

\renewcommand\theequation{\thesection.\arabic{equation}}
\setcounter{equation}{0}

\renewcommand\thefigure{\thesection.\arabic{figure}}    
\setcounter{figure}{0}    

\section{Appendix}
This appendix provides mathematical details for the four discussed flow allocation schemes. For a broader formal treatment, please refer to the original publications that introduce the respective schemes. Throughout this section, let~$\gamma_t = \left( \sum_n \injection^+\right) ^{-1}$ denote the inverse of the total positive injected power~$p^{+}$. 


\subsection{Average Participation}\label{sec:appendix_ap}

\begin{subequations}
Allocating power under the $AP$ allocation scheme is derived from~\citet{achayuthakan_electricity_2010-1}. In a lossless network, the downstream and upstream formulations result in the same peer-to-peer allocation. Therefore, we restrict ourselves to the downstream formulation. In a first step, we define a time-dependent auxiliary matrix~$\InverseAPInjection_t$. It is the inverse of an~$(N\times N)$ matrix that collects the directed power flows~$m \rightarrow n$ at entries~$(m, n)$ for~$m \ne n$, and the total flow passing node~$m$ at entries~$\left( m, m\right)$. Mathematically this translates to
\begin{align}
\InverseAPInjection_t = \left( \diag{\injectionM^+} + \DirectedIncidence^- \diag{\flowM} \, \incidenceM \right)_t^{-1},
\end{align}
where~$\DirectedIncidence_{n,\ell, t} = \text{sign}\left( \flow \right)  \incidence$ represents the directed incidence matrix of which $\DirectedIncidence^-$ collects all negative entries,  $\incidenceM$ is the incidence matrix, and~$\flowM$ the~$(L\times 1)$ vector of all flows~$\flow$. Then, the allocation from source to sink for time step~$t$ is given by
\begin{align}
\allocatePeer = \InverseAPInjection_{n,m,t} \, \netproduction \, \netconsumption[m]
\end{align}

An allocation that considers the total amount of produced energy must also take nodal self-consumption~$\selfconsumption = \text{min}\left( \netproduction, \netconsumption \right)$ into account. Therefore, the peer-to-peer allocation is adjusted to 
\begin{align}
\allocatePeer = \InverseAPInjection_{n,m,t} \, \netproduction \, \netconsumption[m] + \delta_{n,m} \selfconsumption,
\end{align}
where~$\delta_{n,m}$ is zero for~$n\ne m$ and one else. 
Eventually, the flow on line~$\ell$ induced by~$\allocatePeer$ is given by 
\begin{align}
\allocateTransaction = \flow \cdot \left( \sum_{i} \, \DirectedIncidence^+_{i,,\ell}\, \injectionM^+_i \, \InverseAPInjection_{i, n}  \right)_t  \cdot \left( \sum_{i} \DirectedIncidence^-_{i, \ell} \, \injectionM^-_i \, \InverseAPInjection_{m, i}  \right)_t 
\end{align}

The first parentheses specify through which lines power flows that is injected by sources in the network; the second parentheses specify through which lines power flows that is withdrawn by the demand of sinks.

\end{subequations}

\subsection{Marginal Participation}
\label{sec:appendix_mp}

Marginal Participation flow allocation was first introduced by~\citet{rudnick_marginal_1995}. The induced virtual injection pattern~$\allocateInjection$ of node~$n$ for the $MP$ scheme with an equal contribution between consumers and producers is given by 
\begin{align}
\allocateInjection = \delta_{n,m} \injection - \dfrac{1}{2} \gamma_t \, | \injection | \, \injection[m]
\end{align}

The first term is only non-zero if~$n=m$. The second term specifies the share of each node how it contributes to total net injected power in the network. It relates to the proportion to which each node absorbs the infinitesimal increase in power injection of the node~$n$ in question, and thus captures the idea of the marginal participation allocation approach. The flow allocation can be either carried out using the virtual injection patterns~$\allocateInjection$ and equation~\eqref{eq:alloc_vip_to_allocflow} as laid out in this paper or, alternatively, by directly comparing the flow pattern in the infinitesimally altered system to the basic flow pattern without the infinitesimal increase.


\subsection{Equivalent Bilateral Exchanges}\label{sec:appendix_ebe}

The concept of Equivalent Bilateral Exchanges was first introduced by~\citet{galiana_transmission_2003}. To weigh the contribution of producers and consumers, one can introduce a weighting scalar~$q$. The virtual injection pattern is given by 
\begin{align}
\allocateInjection = q\left( \delta_{n,m} \netproduction + \gamma_t \, \netconsumption \, \netproduction[m]  \right) + (1-q) \left(\delta_{n,m} \netconsumption - \gamma_t \, \netproduction\, \netconsumption[m] \right) \label{eq:allocated_injection_ebe}
\end{align}

The first term represents injection patterns of net producers which deliver power to sinks proportionally to their net consumption. The second term comprises injection patterns of all net consumers that retrieve power from sources proportionally to their net production. Throughout the paper, the weighting scalar is set to the default value of~$q=0.5$. 


\subsection{Linearized Z-bus allocation}\label{sec:appendix_zbus}

\newcommand{\linevoltage}{v_{\ell,t}}
\newcommand{\zbus}[1][n,m]{Z_{#1}}
\newcommand{\cisf}{H_{n,\ell}}
\newcommand{\voltage}{\B{v}}
\newcommand{\current}{i_{n,t}}
\newcommand{\admittance}{y_\ell}
\newcommand{\re}{\operatorname{Re}}

\begin{subequations}
The \textit{Z-bus} transmission allocation presented in~\citet{conejo_z-bus_2007} is based on circuit theory. Taking the full AC power flow equations, the authors define an electrical distance from which the contribution of each power source and sink is derived. Let~$()^*$ denote the complex conjugate. The original \textit{Z-bus} allocation is given by 
\begin{align}
\allocateFlow = \re\left( \linevoltage \left( \admittance \sum_m \incidence[m] \, \zbus[m,n]  \right) ^* \, \current^*\right), \label{eq:Zbus_allocation}
\end{align} 
where~$\linevoltage$ is the voltage level on line~$\ell$, averaged over its starting and end point, $\admittance$ its admittance, $\zbus$ the standard Z-bus matrix, and~$\current$ the nodal current injection. Note that this formulation contains an implicit fifty-fifty weighting because it does not differentiate between sources and sinks. In the linear power flow approximation, the terms in the inner parentheses translate to the PTDF
\begin{align}
\admittance \sum_m \incidence[m] \, \zbus[m,n] \xrightarrow[]{\text{linear approx.}} \ptdf
\end{align}
Further, as voltage differences between to adjacent nodes are neglected, we obtain that 
\begin{align}
\allocateFlow = \ptdf \, \injection
\end{align}

\end{subequations}


\subsection{Network for the working example}\label{sec:appendix_network}

\begin{figure}[h]
\centering 
\includegraphics[width=1.0\linewidth]{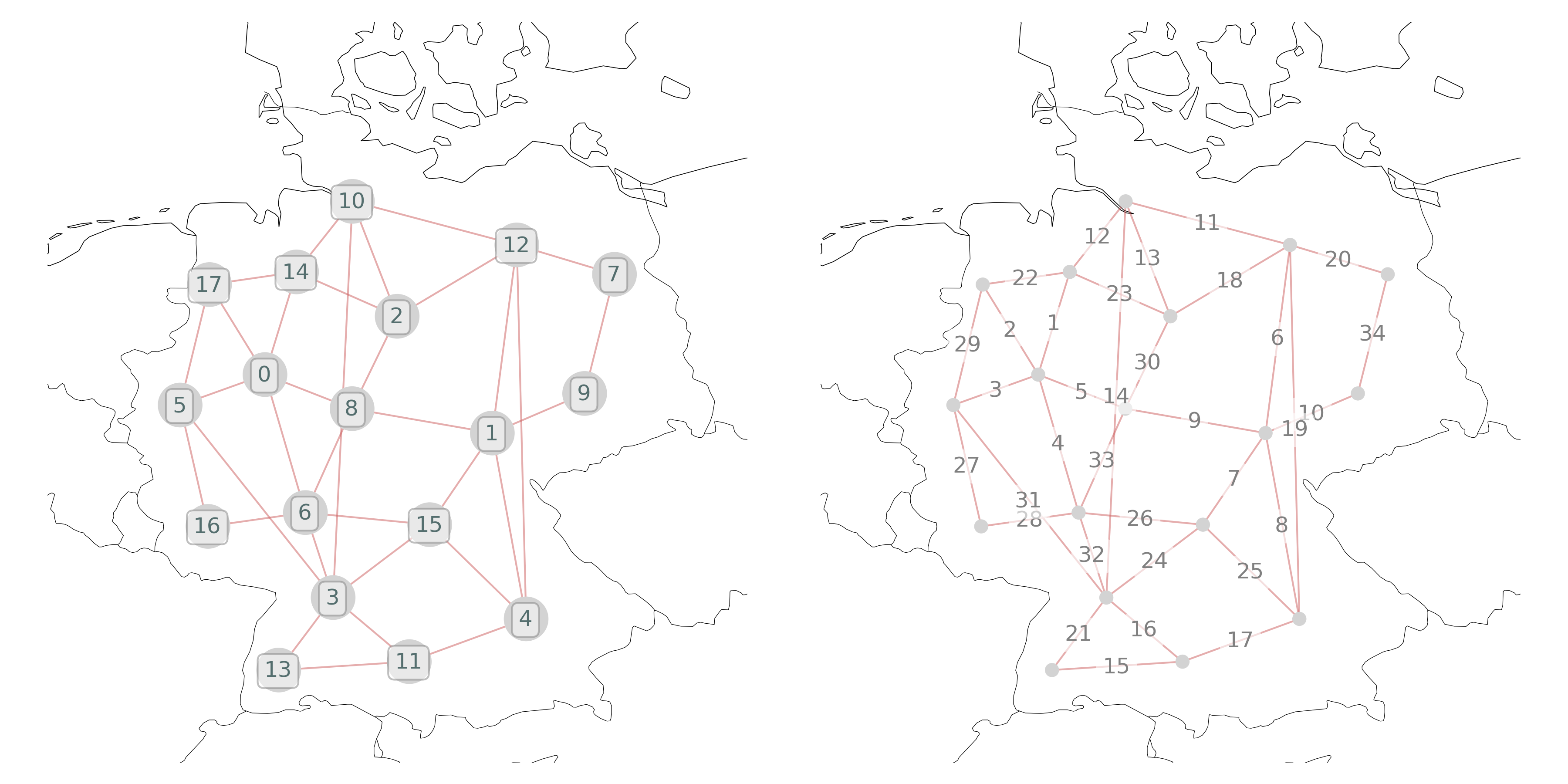}
\caption{Topology of the network for the working example}
\label{fig:app_topology}
\end{figure}

Figure~\ref{fig:app_topology} illustrates the topology of our German sample network including the denotations for nodes and lines.

%


\end{document}